\documentclass[a4paper,11pt]{article}
\usepackage{epsfig,amsmath,amssymb,amsthm}
\usepackage{openbib}
\usepackage[round]{natbib}
\usepackage[ansinew]{inputenc}% Richtige Umlaute
\usepackage{graphicx}
\usepackage{enumerate}
\usepackage{lscape}
\usepackage{fullpage}
\usepackage{setspace}
\usepackage{enumitem}
\usepackage{subfigure}
\usepackage{rotating}

\relax

\newcommand{\mathp}{\mathsf{P}}

\usepackage{verbatim,color}

\def\boxit#1{\vbox{\hrule\hbox{\vrule\kern6pt
          \vbox{\kern6pt#1\kern6pt}\kern6pt\vrule}\hrule}}

\DeclareMathOperator*{\argmax}{argmax}

\newcommand{\captionfonts}{\small}

\makeatletter  % Allow the use of @ in command names
\long\def\@makecaption#1#2{%
  \vskip\abovecaptionskip
  \sbox\@tempboxa{{\captionfonts #1: #2}}%
  \ifdim \wd\@tempboxa >\hsize
    {\captionfonts #1: #2\par}
  \else
    \hbox to\hsize{\hfil\box\@tempboxa\hfil}%
  \fi
  \vskip\belowcaptionskip}
\makeatother   % Cancel the effect of \makeatletter
\title{\textsc{Multiple break detection in the correlation structure of random variables}}

\author{\textsc{Pedro Galeano, Dominik Wied}\footnote{Universidad Carlos III de Madrid, Departamento de Estadística, E-28903 Getafe, Madrid, Spain. Email: pedro.galeano@uc3m.es, Phone: +34 91 624 8901 (P. Galeano). TU Dortmund, Fakultät Statistik, D-44221 Dortmund, Germany.\ Email: wied@statistik.tu-dortmund.de, Phone: +49 231 755 3869 (D. Wied).} \\ \small \textit{Universidad Carlos III de Madrid and TU Dortmund}}

\date{This Version: \today}

\begin{document}

%\doublespacing

\parindent 0cm

\makeatletter
\def\@seccntformat#1{\csname the#1\endcsname. }
\def\section{\@startsection {section}{1}{\z@}{-3.5ex plus -1ex minus
    -.2ex}{1.3ex plus .2ex}{\center\large\sc}}
\def\subsection{\@startsection{subsection}{2}{\z@}{3.25ex plus 1ex minus .2ex}{-1em}{\normalsize\bf}}

\def\subsubsection{\@startsection{subsubsection}{3}{\z@}{3.25ex plus 1ex minus .2ex}{-1em}{\normalsize\it}}

\newtheorem{theorem}{Theorem}
\newtheorem{definition}{Definition}
\newtheorem{lemma}{Lemma}
\newtheorem{assumption}{Assumption}
\newtheorem{cor}{Corollary}
\theoremstyle{definition}
\newtheorem{example}{Example}
\newtheorem{remark}{Remark}

\bibliographystyle{ecta}

%===Deckblatt=======================================================%

\maketitle
%===================================================================%

\begin{abstract}
Correlations between random variables play an important role in applications, e.g.\ in financial analysis. More precisely, accurate estimates of the correlation between financial returns are crucial in portfolio management. In particular, in periods of financial crisis, extreme movements in asset prices are found to be more highly correlated than small movements. It is precisely under these conditions that investors are extremely concerned about changes on correlations. A binary segmentation procedure to detect the number and position of multiple change points in the correlation structure of random variables is proposed. The procedure assumes that expectations and variances are constant and that there are sudden shifts in the correlations. It is shown analytically that the proposed algorithm asymptotically gives the correct number of change points and the change points are consistently estimated. It is also shown by simulation studies and by an empirical application that the algorithm yields reasonable results.
\end{abstract}


%\textbf{JEL Classification:} C12, C14, C63, G12

\section{Introduction and Summary} \label{sec:intro}

There is much empirical evidence that the correlation structure of financial returns of all sorts cannot be assumed to be constant over time, see
e.g. \citet{krishan:2009}. Especially in times of crisis, correlation often increases, a phenomenon which is referred to as ``Diversification
Meltdown" (\citealp{campbell:2008}). Wied, Krämer and Dehling (2012)\nocite{wied:2011} propose a CUSUM type procedure along the lines of
\citet{ploberger:1989} to formally test if correlations between random variables remain constant over time. However, with this approach the
practitioner is only able to see if there is a change or not; he cannot determine where a possible change occurs or how many changes there are.

The present paper fills this gap by proposing an algorithm based on the correlation constancy test to estimate both the number and the timing of
possible change points. For this purpose, we adapt a method for the estimation of multiple breaks from \citet{vostrikova:1981} which has been
implemented in various problems by \citet{inclan:1994}, \citet{bai:1997}, \citet{bai:1998}, \citet{andreou:2002}, \citet{degooijer:2006}, \citet{galeano:2007b} and
\citet{galeano:2010a}, among others. The segmentation algorithm proceeds as follows: First, we determine the ``dominating'' change point and decide
if this point is statistically significant. Then, we split the series in two parts and again test for possible change points in each part of the
series. The procedure stops if we do not find any new change point any more. In this paper, we will analytically show that the algorithm
asymptotically gives the correct number of change points and that - finitely many - change points are consistently estimated. Furthermore, we show
that the algorithm gives reasonable results in finite samples and in an empirical application.

The rest of the paper is organized as follows. Section 2 introduces the proposed procedure. Section 3 derives the asymptotic properties of the
procedure. Sections 4 and 5 present some simulation studies and a real data application and Section 6 provides some conclusions. All proofs are
presented in the Appendix.

\section{An algorithm for the determination of change points}
\label{sec:algorithm}

In this section, we present the algorithm for the detection of change points in the correlation structure of bivariate random variables. To be more precise, let $(X_t,Y_t), t \in \mathbb{Z},$ be a sequence of bivariate random variables with finite first four moments and let $1,\ldots,T$ be the observation period. Denoting the correlation between $X_t$ and $Y_t$ by
\begin{align*}
\rho_t = \frac{Cov(X_t,Y_t)}{\sqrt{Var(X_t)}\sqrt{Var(Y_t)}},
\end{align*}
\citet{wied:2011} propose a test for the problem
\begin{align*}
H_0: \rho_1 = \ldots = \rho_T \text{ vs. } H_1: \exists t \in \{1,\ldots,T-1\}: \rho_t \neq \rho_{t+1}
\end{align*}
which uses the test statistic
\begin{equation}
Q_T(X,Y) = \hat D \max_{2 \leq j \leq T} \frac{j}{\sqrt{T}} \left| \hat \rho_j - \hat \rho_T \right|, \label{Q_T}
\end{equation}
where $\hat \rho_j$ is the empirical correlation up to time $j$, for $j=2,\ldots,T$ and $\hat D$ is a normalizing constant which is described in Appendix A.1. \citet{wied:2011} show that the asymptotic null distribution of $Q_T(X,Y)$ is the supremum over the absolute value of a standard
Brownian bridge. The present paper employs this test to estimate the timings and the number of possible change points.

We assume that there is a finite number of change points. However, the number, location and size of the change points are unknown. \citet{wied:2011} allow for some fluctuations in the first and second moments under the null hypothesis, compare their assumption (A4). The variance fluctuations may be slightly stronger if the variances behave similarly, but it has to be stressed that no arbitrary fluctuations are allowed. In fact, in both settings, the variance shifts vanish with increasing sample size. For this reason and for ease of exposition, we focus on the case where expectations and variances are constant under the alternative. It would be possible to extend the framework to slightly changing expectations and variances as described in the previous paragraph, but this would only affect the proofs, not the procedure itself. Note that stationary GARCH models are included in our setup as the unconditional variances are constant here. We investigate in our simulation study how the procedure behaves in finite samples in the presence of GARCH effects (volatility clustering) or shifts in the mean.

% which are rather irrelevant for financial return data though) and in the case in which a pre-filtering is performed.

The formal assumption is:

\begin{assumption}\label{ass:MomentsFunctiong}
Under the alternative, expectations and variances are constant and equal to finite numbers $\mu_x,\mu_y,\sigma_x^2$ and $\sigma_y^2$,
the second cross moment changes from $\mathsf{E}(X_t Y_t) = m_{xy}$ to $\mathsf{E}(X_t Y_t) = m_{xy} + g\left(\frac{t}{T} \right)$. The function $g(z), z \in [0,1]$ is a step function with a finite number of steps $\ell$, i.e. there is a partition $0 = z_0 < z_1 < \ldots < z_{\ell} < z_{\ell+1} = 1$ and there are second cross moment levels $a_0,\ldots,a_{\ell}$ such that
\begin{equation*}
g(z) = \sum_{i=0}^{\ell} a_i {\bf 1}_{\{z \in [z_i,z_{i+1})\}}
\end{equation*}
and $g(1) = a_{l}$. The quantities $\ell$, $z_1,\ldots,z_{\ell}$ and $a_0,\ldots,a_{\ell}$ do not depend on $T$.
\end{assumption}

The function $g$ specifies the timing and the size of the changes in correlation. Since this is a step function, we consider sudden changes in the correlation (or more specific the covariance) and do not consider smooth changes.

Our goal is to estimate $\ell$, $z_1,\ldots,z_{\ell}$ and $a_0,\ldots,a_{\ell}$. To this end, we propose a binary segmentation algorithm.
The main idea is to isolate each change point in different time intervals by splitting the two series into two parts once a change point is found.
Then, the search of a new change point is repeated in both sections.
The proposed procedure for detecting correlation changes essentially relies on the intuitive estimator of the change point fraction.
To that purpose, we rewrite the test statistic \eqref{Q_T} as
\begin{equation*}
Q_T(X,Y) = \sup_{z \in [0,1]} \hat D \frac{\tau(z)}{\sqrt{T}} \left| \hat \rho_{\tau(z)} - \hat \rho_T \right|
\end{equation*}
with $\tau(z) = [2+z(T-2)]$ (where $[\cdot]$ is the floor function) and estimate the timing of the break by $\hat z := \tau(\hat z^*)/T$ with $\hat z^* := \mathsf{argmax}_z B_{T}(z)$ and
$B_T(z) := \hat D \frac{\tau(z)}{T} \left| \hat \rho_{\tau(z)} - \hat \rho_T \right|$.
Here and in the following, we restrict the values $z$ for which the $\mathsf{argmax}$ is calculated to multiples of $1/T$ and, for uniqueness, then choose the smallest of these values.
Note that $B_T(z)$ is calculated from all observations. In the next steps of the algorithm, we just consider the observations in the relevant part of the sample and we call the corresponding ``target function'' $|A_T(z)|$, where, for $0 \leq l_1 < l_2 \leq 1$ and $z \in [l_1,l_2]$,
\begin{equation*}
A_{T}(z):=\hat D \frac{\xi(z) - \eta(l_1)+1}{\xi(l_2) - \eta(l_1)+1} \left( \hat \rho_{\eta(l_1)}^{\xi(z)} - \hat \rho_{\eta(l_1)}^{\xi(l_2)} \right).
\end{equation*}
Here, $\eta(z) = ([zT] \vee 1) \wedge (T-1)$, $\xi(z) = \eta(z) \vee (\eta(l_1)+1)$ (where $\vee$ and $\wedge$ stands for maximum and minimum, respectively) and $\hat \rho_a^b$ denotes the empirical correlation coefficient calculated from data point $a$ to data point $b$. Moreover, $\hat D$ is the variance estimator from Appendix A.1 calculated from the data from $\eta(l_1)$ to $\xi(l_2)$. Then the timing of break is estimated by

\begin{equation}\label{est}
\hat z := \xi(\hat z^*)/T
\end{equation}
with $\hat z^* = \mathsf{argmax}_{l_1 \leq z \leq l_2} |A_T(z)|$.

Basically, this means that we always look for the time point at which the test statistic \eqref{Q_T} (calculated from data in a particular interval) takes its maximum and divide by $T$. Note that $B_T(z) = A_T(z)$ for $l_1=0$ and $l_2=1$.

The formal algorithm proceeds as follows:\bigskip

\begin{enumerate}
\item Let $X_{t}$ and $Y_{t}$ be the observed series. Obtain the test statistic $Q_T(X,Y)$. There are two possibilities:

\begin{enumerate}
\item If the test statistic is statistically significant, i.e., if $Q_T(X,Y)>c_{T,\alpha }$, where $c_{T,\alpha }$ is the asymptotic critical
value for a given upper tail probability, then a correlation change is announced. Let $z_{1}$ be the break point estimator from \eqref{est} and go to step 2.

\item If the test statistic is not statistically significant, the algorithm stops.
\end{enumerate}

\item Let $z_{1},\ldots ,z_{\ell }$ be the $\ell $ change points in increasing order already found in previous iterations. Repeat step 1 for every segment until
\begin{equation*}
\underset{k}{\max }\left\{ Q_{T}^{k}(X,Y),\text{ }k=1,\ldots ,\ell +1\right\}
<c_{T,\alpha },
\end{equation*}
where $Q_{T}^{k}(X,Y)$ is the value of the statistic $Q_T(X,Y)$ calculated from the data from $\eta(z_{k-1}+1/T)$ to $\xi(z_k)$, for $k=1,\ldots ,\ell +1$, taking $z_{0}=0$ and $z_{\ell +1}=1$.

\item Let $\left( z_{1} < \ldots < z_{\ell }\right) $ be the detected change points. If $\ell >1$, refine the
estimate of the location of the change points by calculating the statistic $Q_T(X,Y)$ from the data from $\eta(z_{k-1}+1/T)$ to $\xi(z_{k+1})$, for $k=1,\ldots ,\ell $, where $z_{0}=0$ and $z_{\ell +1}=1$. If any of the change points is not statistically
significant, delete it from the list, and repeat this step.

\item Finally, estimate the correlation between $X_{t}$ and $Y_{t}$ in each segment separately with the usual Bravais-Pearson correlation coefficient.
\end{enumerate}

\bigskip

The key point of the proposed procedure is that it detects a single change point in each iteration, which may not be the most efficient way to
detect correlation changes when multiple changes exist. However, our theoretical results show that the procedure consistently detects the true
change points. Moreover, the proposed procedure works well in small samples in terms of detection of the true number of changes as shown in the
Monte Carlo experiments of Section 4.

Step 3 is meant to refine the estimation of the change points. Note that in this step, the procedure computes the value of the $Q_T(X,Y)$ statistics in intervals that are only affected by the presence of a single change point, which is not guaranteed in step 2.

In a sense, the main objective of the proposed procedure is to identify issues which require further attention. For instance, if the number of
change points detected is large compared to the sample size, then a piecewise constant correlation may not be a good description of the true
correlation between the two series.

Although we later prove that we can consistently estimate the correct number of change points even if the critical value is the same in each step of the procedure, we proceed differently in practice with finite samples. Using the same critical level in steps 2 and 3 may lead to over-estimation of the number of change points, because more tests are performed in each iteration as the number of detected change points increases and the type I errors accumulate (remember that the decision of the tests basically determine the number of change points). So, to avoid this multiple-test problem we require that the type I errors used depend on the number of change points already detected by the algorithm. To be more precisely, if $\alpha_0$ is a fixed initial type I error for step 1 such as $\alpha_0 = 0.05$, we use the critical value $c_{T,\alpha_k}$ after detecting the $(k-1)$-th change point. Here, $\alpha_k$ is such that $1-\alpha _{0}=(1-\alpha_k)^{k+1}$. This leads to $\alpha_k = 1 - (1-\alpha_0)^{\frac{1}{k+1}}$, so for example to $\alpha_1 \approx 0.025$ and $\alpha_2 \approx 0.017$ for $\alpha_0 = 0.05$. This choice of $\alpha_k$ keeps the same significance level constant for all tests. In fact, for the asymptotic result concerning the number of break points (Theorem 2), the initial type I error would have to converge to zero, but in finite samples, $\alpha_0 = 0.05$ seems to be an acceptable choice. Moreover, in practice, we use the quantiles of the distribution of the supremum of the absolute value of a standard Brownian bridge (the limit distribution of the correlation test statistic under the null hypothesis) in order to apply the procedure. The explicit form of this distribution function can be found in \citet{billingsley:1968}, p. 85. For example, for $\alpha_{0}=0.05$ the critical value is $1.358$.

\section{Asymptotic results}\label{sec:Asymp}

In this section, we show that our algorithm asymptotically gives correct solutions. To this end, we impose another assumption which guarantees that we do not have two or more change points with ``equal form'', i.e.\ we assume that there are always change points which dominate the rest.

\begin{assumption}\label{ass:dominatingChangePoint}
Let $0 \leq l_1 < l_2 \leq 1$ be arbitrary. The function $g$ from Assumption \ref{ass:MomentsFunctiong} is such that the function $|A^*(z)|$ with
\begin{equation*}
A^*(z) := \int_{l_1}^z g(t) dt - \frac{z - l_1}{l_2 - l_1} \int_{l_1}^{l_2} g(t) dt, z \in [l_1,l_2],
\end{equation*}
is either constant or has a unique maximum.
\end{assumption}

A dominating change point is then defined as $\mathsf{argmax}_z |A^*(z)|$ in a given interval $[l_1,l_2]$. For ease of exposition, we do not mention the dependence of $A^*(z)$ from $l_1$ and $l_2$ which should become clear from the context. We illustrate Assumption \ref{ass:dominatingChangePoint} in the case of the example

\begin{equation}\label{ExampleG1}
g(z) = 0.5 \cdot {\bf 1}_{\{z \in [0,0.5)\}} + 0.7 \cdot {\bf 1}_{\{z \in [0.5,0.75)\}} + 0.6 \cdot {\bf 1}_{\{z \in [0.75,1]\}}.
\end{equation}

\begin{figure}[!tbp]
\centering
\caption{Function $A^*(z)$ in example \eqref{ExampleG1} for $z \in [0,1]$} \label{Figure1}
\includegraphics[width=6.0in,height=3.5in]{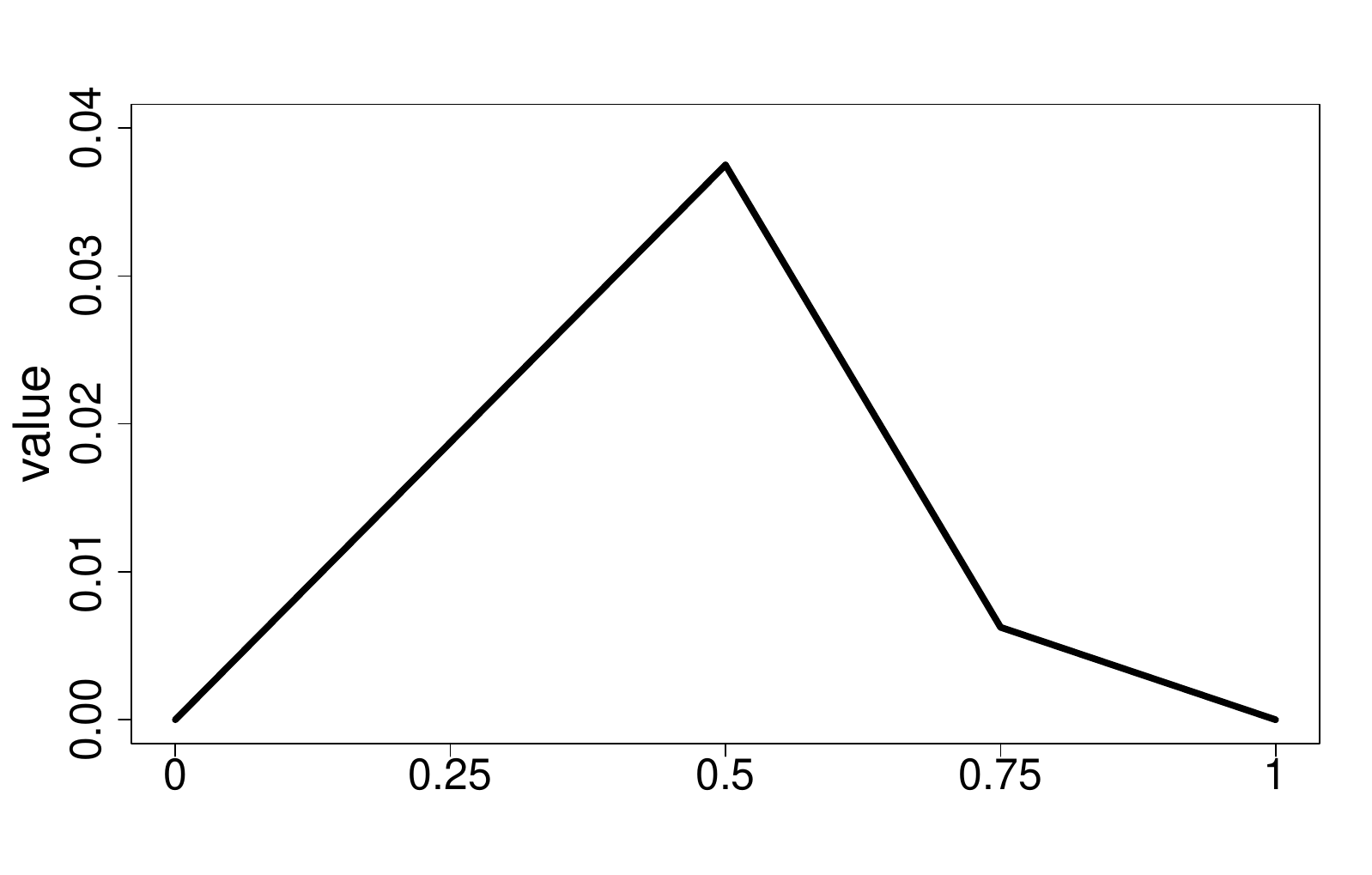}
\end{figure}

In the interval $[0,1]$ for example, we have
\begin{equation*}
|A^*(z)| = \begin{cases} 0.075 \cdot z & z < 0.5, \\ 0.1 - 0.125 \cdot z & 0.5 \leq z < 0.75, \\ 0.025 - 0.025 \cdot z & 0.75 \leq z \leq 1 \end{cases},
\end{equation*}
so $|A^*(z)|$ has a unique maximum at $z = 0.5$, i.e. the point with the ``strongest'' correlation change, see also Figure \ref{Figure1}.

\bigskip

\begin{center}
\bf{Figure \ref{Figure1} about here}
\end{center}

\bigskip

\begin{figure}[!tbp]
\centering
\caption{Function $A^*(z)$ in example \eqref{ExampleG2} for $z \in [0,1]$} \label{Figure2}
\includegraphics[width=6.0in,height=3.5in]{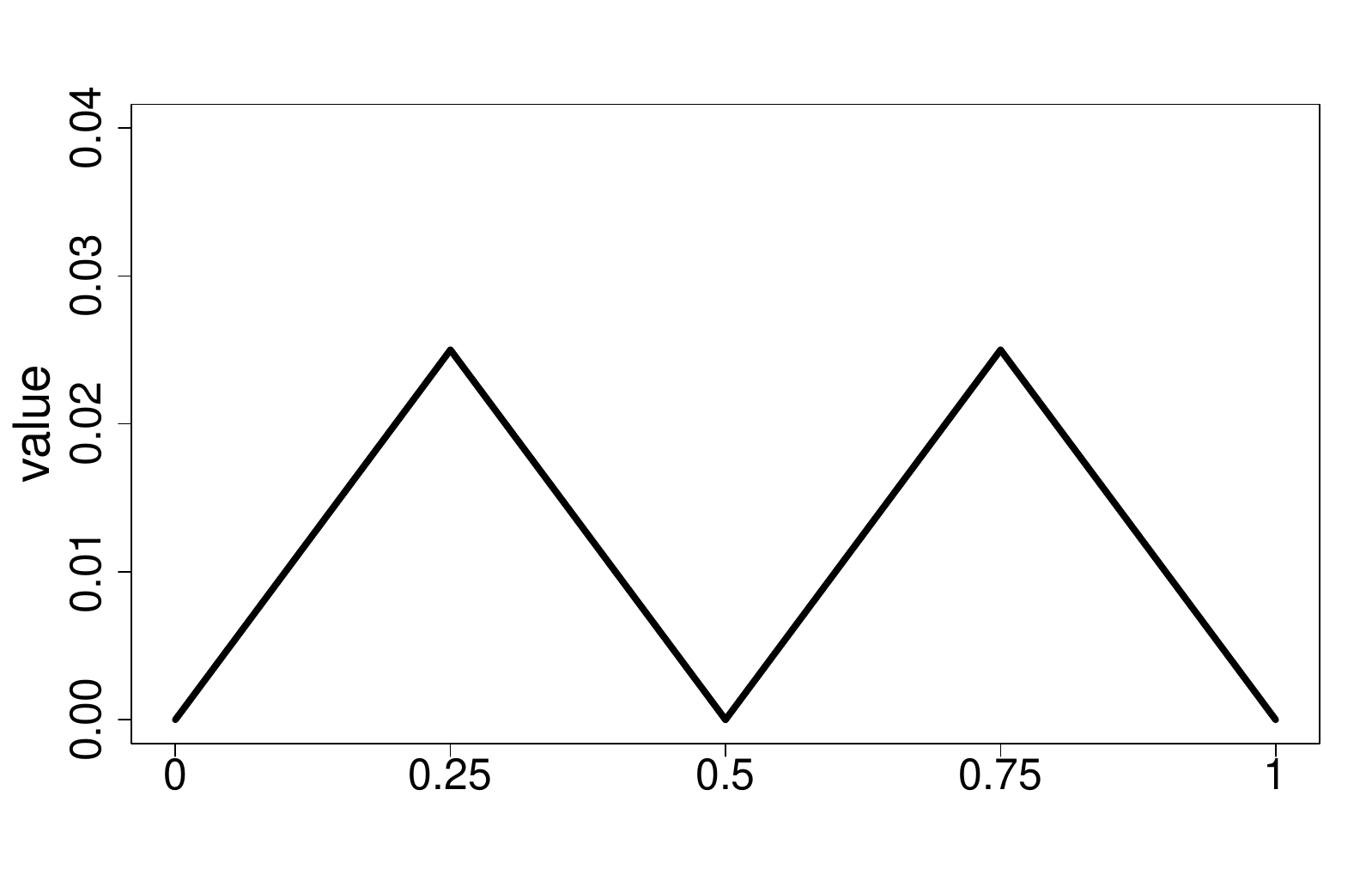}
\end{figure}

In general, the size and the position of the change decide if it is dominant or not. Assumption \ref{ass:dominatingChangePoint} is violated if the correlation changes are equal at symmetric time points, e.g. when
\begin{equation}\label{ExampleG2}
g(z) = 0.5 \cdot {\bf 1}_{\{z \in [0,0.25)\}} + 0.7 \cdot {\bf 1}_{\{z \in [0.25,0.75)\}} + 0.5 \cdot {\bf 1}_{\{z \in [0.75,1]\}}.
\end{equation}
Here, we have
\begin{equation*}
|A^*(z)| = \begin{cases} 0.1 \cdot z & \text{ when } z < 0.25, \\ 0.05 - 0.1 \cdot z & \text{ when } 0.25 \leq z < 0.5, \\ 0.1 \cdot z - 0.05 & \text{ when } 0.5 \leq z < 0.75, \\ 0.1 - 0.1 \cdot z & \text{ when } 0.75 \leq z \leq 1 \end{cases},
\end{equation*}
so $|A^*(z)|$ has two non-unique maxima at $z=0.25$ and $z=0.75$, see also Figure \ref{Figure2}.

\bigskip

\begin{center}
\bf{Figure \ref{Figure2} about here}
\end{center}

\bigskip

Moreover, we need a rather technical assumption regarding the normalizing constant $\hat D$.

\begin{assumption}\label{ass:dhatAlternative}
Consider an arbitrary interval $[l_{1},l_{2}]\subseteq \lbrack 0,1]$ with $l_1 < l_2$ and let $\hat D$ be the estimator from Appendix A.1 calculated from data from $\eta(l_1)$ to $\xi(l_2)$.
Then, under the alternative, $\hat D$ converges to a real number $D_1^A \in (0,\infty)$.
\end{assumption}

For the interval $[0,1]$, this assumption is for example fulfilled in a simple model with a bivariate normal distribution, serial independence,
constant expectations $0$, constant variances $1$ and a change in the covariance in the middle of the sample from $\rho_1$ to $\rho_2$.
In this case, it is easy to see that $D_1$ is equal to $(D_c D_b D_a D_b^{'} D_c^{'})^{-1/2}$ with $$D_a = \begin{pmatrix} 2 & \rho_{1,2} & 0 & 0 & \rho_{1,2} \\ \rho_{1,2} & 2 & 0 & 0 & \rho_{1,2} \\ 0 & 0 & 1 & \frac{\rho_{1,2}}{2} & 0 \\ 0 & 0 & \frac{\rho_{1,2}}{2} & 1 & 0 \\ \rho_{1,2} & \rho_{1,2} & 0 & 0 & 1 \end{pmatrix}\ D_b = \begin{pmatrix} 1 & 0 & 0 & 0 & 0 \\ 0 & 1 & 0 & 0 & 0 \\ 0 & 0 & 0 & 0 & 1 \end{pmatrix},$$ $D_c = (-\frac{\rho_{1,2}}{4} , -\frac{\rho_{1,2}}{4} , 1 )$ and $\rho_{1,2} = \rho_1+\rho_2$. So, $D_1$ is equal to $\left( 1 - \frac{3}{4} \cdot \rho_{1,2}^2 + \frac{1}{8} \cdot \rho_{1,2}^3 \right)^{-1/2}$ and this is a real positive number e.g. for $\rho_1 = - \rho_2$.

Moreover, we need two assumptions concerning bounded moments and serial dependence.

\begin{assumption}\label{boundedmoments1}
The $q$-th absolute moments of the components of $U_t = (X_t^2,Y_t^2,X_t,Y_t,X_t Y_t)$ are uniformly bounded for some $q > 1$.
\end{assumption}

\begin{assumption}\label{serialdependence1}
The vector $(X_t,Y_t)$ is $L_p$-NED (near-epoch dependent) with size $-1/p$, where $1 < p \leq 2$ and $p \leq q$ with $q$ from Assumption \ref{boundedmoments1}, and constants $(c_t),t \in \mathbb{Z}$, on a sequence $(V_t), t \in \mathbb{Z}$, which is $\alpha$-mixing of size $\phi^* :=-r/(r -2)$ with $2 < r < 2 q$, i.e.,
\begin{equation*}
||(X_t,Y_t) - \mathsf{E}((X_t,Y_t)|\sigma(V_{t-m},\ldots,V_{t+m}))||_p \leq c_t v_m
\end{equation*}
with $\lim_{m \rightarrow \infty} v_m = 0$, such that $c_t \leq 2 ||U_t||_p$ with $U_t$ from Assumption \ref{boundedmoments1} and the $L_p$-norm $|| \cdot ||_p$.
\end{assumption}

The following theorem shows that the change point estimator \eqref{est} is consistent if it is known a priori that there is a change point in a given interval.

\begin{theorem}\label{theorem:ConsistencySeveralBreaks}
Let Assumptions \ref{ass:MomentsFunctiong}, \ref{ass:dominatingChangePoint}, \ref{ass:dhatAlternative}, \ref{boundedmoments1} and \ref{serialdependence1} be true and let there be at least one break point in a given interval $[l_{1},l_{2}]\subseteq \lbrack 0,1]$ with $l_1 < l_2$. Then the change point estimator \eqref{est} is consistent for the dominating change point.
\end{theorem}

While also of interest on its own, this theorem is mainly needed for the next one which provides the convergence of the algorithm. Note that, as we just use a law of large numbers and no (functional) central limit theorem in the proof of this theorem, we just need a little bit more than finite second moments. This is different in the following theorem yielding consistency of the number of change points. Here, we need stronger assumptions to guarantee that the test statistic behaves well in the case of no correlation change. These assumptions are basically similar to the assumptions (A1), (A2) and (A3) in \citet{wied:2011}.

\begin{assumption}\label{ass:dhatAlternative2}
Consider an arbitrary interval $[l_{1},l_{2}]\subseteq \lbrack 0,1]$ with $l_1 < l_2$. For $S_j := \sum_{t = 1 \vee [l_1 T]}^{j} (U_t - \mathsf{E}(U_t))$ it holds under the null $\lim_{T \rightarrow \infty} \frac{1}{(l_2-l_1)T} \mathsf{E} S_{[l_2 T]} S'_{[l_2 T]} = D_N$ which is a finite and positively definite matrix.
\end{assumption}

\begin{assumption}\label{boundedmoments2}
The $r$-th absolute moments of the components of $U_t = (X_t^2,Y_t^2,X_t,Y_t,X_t Y_t)$ are uniformly bounded for some $r > 2$.
\end{assumption}

\begin{assumption}\label{serialdependence2}
The vector $(X_t,Y_t)$ is $L_2$-NED (near-epoch dependent) with size $-(r-1)/(r -2)$, where $r$ is from Assumption \ref{boundedmoments2}, and constants $(c_t),t \in \mathbb{Z}$, on a sequence $(V_t), t \in \mathbb{Z}$, which is $\alpha$-mixing of size $\phi^* :=-r/(r -2)$, i.e.,
\begin{equation*}
||(X_t,Y_t) - \mathsf{E}((X_t,Y_t)|\sigma(V_{t-m},\ldots,V_{t+m}))||_2 \leq c_t v_m
\end{equation*}
with $\lim_{m \rightarrow \infty} v_m = 0$, such that $c_t \leq 2 ||U_t||_2$ with $U_t$ from Assumption \ref{boundedmoments2} and the $L_2$-norm $|| \cdot ||_2$.
\end{assumption}

Assumption \ref{ass:dhatAlternative2} is for example fulfilled in the situation described after Assumption \ref{ass:dhatAlternative} with $\rho_1 = \rho_2$.

The following theorem also requires an additional assumption on the critical values $c_{T,\alpha_k}$. While we argued in the preceding section that we have to adjust the value for finite $T$ due to multiple testing problems, we need another kind of assumption for the asymptotics as $T \rightarrow \infty$. This assumption is basically the same as in \citet{bai:1997}, Proposition 11.

\begin{assumption}\label{ass:CriticalValue}
The critical values $c_{T,\alpha_k}$ used in the algorithm obey the condition \\ $\lim_{T \rightarrow \infty} c_{T,\alpha_k} = \infty$ and $c_{T,\alpha_k} = o(\sqrt{T})$ for $k \in \mathbb{N}_0$.
\end{assumption}

The assumption rules out choosing an initial type I error such as $\alpha_0=0.05$ for all $T \in \mathbb{N}$ because the initial type I error must converge to $0$. However, it is legitimate using a fixed type I error in finite samples if we consider an upper bound for $T$. The initial type I
error $\alpha_0=0.05$ then corresponds to $c_{T,\alpha_0} = 1.358$, see the discussion before Section 3. Note that the level adjustment discussed in the previous section basically fits into this setting because $\ell$ is a fixed number not depending on $T$. This guarantees that the critical values do not become too large so that the $o(\sqrt{T})$-condition is not violated.

\begin{theorem}\label{theorem:AlgorithmWorks}
Under Assumptions \ref{ass:MomentsFunctiong}, \ref{ass:dominatingChangePoint}, \ref{ass:dhatAlternative}, \ref{ass:dhatAlternative2}, \ref{boundedmoments2}, \ref{serialdependence2} and \ref{ass:CriticalValue} the change point algorithm asymptotically gives the correct number of change points $\ell$ and the change points are consistently estimated.
\end{theorem}

Finally, in this section, we want to address the case in which the correlation shifts tend to zero with rate $\frac{1}{\sqrt{T}}$ as the sample size increases such that in Assumption \ref{ass:MomentsFunctiong} we replace $\mathsf{E}(X_t Y_t) = m_{xy} + g\left(\frac{t}{T} \right)$ by $\mathsf{E}(X_t Y_t) = m_{xy} + \frac{1}{\sqrt{T}} g\left(\frac{t}{T} \right)$. In this setting, \citet{wied:2011} provide local power results (compare their Theorem 2). We do not have consistency to the true break point any more, but the change point estimator converges to a non-degenerated random variable as the next theorem shows.

%While in this case, Theorem 1 still goes through by considering $\mathsf{argmax}_z \sqrt{T} B_{T}(z)$ instead of $\mathsf{argmax}_z B_{T}(z)$, it is probably not possible to simply adapt Theorem 2. Given the local power results in \citet{wied:2011}, $Q_T(X,Y)$ converges to a non-degenerate random variable and it does not hold that $\frac{Q_T(X,Y)}{a_T} \rightarrow \infty$ for $a_T = o(\sqrt{T})$ under the alternative. This issue could be addressed in further research.

%However, in this local power setting we can as well consider the asymptotic distribution of our change point estimators, which is necessary to construct asymptotic valid confidence intervals. This follows in the next theorem. It turns out that we get $\sqrt{T}$-convergence to the argmax of a shifted Brownian bridge.

\begin{theorem}\label{theorem:AsyDis2}
Let Assumptions \ref{ass:MomentsFunctiong} (with $\mathsf{E}(X_t Y_t) = m_{xy} + g\left(\frac{t}{T} \right)$ replaced by $\mathsf{E}(X_t Y_t) = m_{xy} + \frac{1}{\sqrt{T}} g\left(\frac{t}{T} \right)$),
\ref{ass:dhatAlternative2}, \ref{boundedmoments2} and \ref{serialdependence2} be true and let there be at least one break point in a given interval $[l_{1},l_{2}]\subseteq \lbrack 0,1]$ with $l_1 < l_2$.
Then it holds for the change point estimator \eqref{est} that
\begin{equation*}
\hat z \rightarrow_d \argmax_{l_1 \leq z \leq l_2} \left| W(z) - W(l_1) - \frac{z - l_1}{l_2 - l_1} \left( W(l_2) - W(l_1) \right) + D_A A^*(z) \right|,
\end{equation*}
where $D_A$ is a constant depending on the data generating process formally defined in the proof, $A^*(z)$ is from Assumption \ref{ass:dominatingChangePoint} and $W(z)$ is a one-dimensional standard Brownian motion.
\end{theorem}

Note that under our local alternatives, the limit matrix from Assumption \ref{ass:dhatAlternative2} is equal to the corresponding limit matrix under the null. Second, note that, for $l_1=0$ and $l_2=1$, the quantity
\begin{equation*}
W(z) - W(l_1) - \frac{z - l_1}{l_2 - l_1} \left( W(l_2) - W(l_1) \right)
\end{equation*}
is a one-dimensional standard Brownian bridge. Third, note that, although the quantity $A^*(z)$ appears in the limit process, we do not need Assumption \ref{ass:dominatingChangePoint} for this theorem. In fact, the (formerly assumed to be existing) unique maximum of $A^*(z)$ does not appear in the limit random variable from Theorem \ref{theorem:AsyDis2}.

\section{Monte Carlo experiments}

\label{sec:finitesample}

This section presents several Monte Carlo experiments to gain insight into the finite sample performance of the proposed procedure. We study several aspects, including the size (probability of a type I error) of the procedure,
its power in correct detection of the changes and its ability to accurately identify the location of the change points. For that, we consider two different scenarios. The first one based on a vector autoregression (VAR) model,
which is widely used for many economic time series, and the second one based on a dynamic conditional correlation (DCC) model, which is widely used for financial returns.

For the first scenario, initially we check the size (probability of a type I error) of the procedure which can be equivalently considered as the accuracy of the estimator of the number of change points if the true value is zero. To this purpose, we consider a vector autoregression of order 1 given by:
\begin{equation*}
\left(
\begin{tabular}{l}
$X_{t}$ \\
$Y_{t}$%
\end{tabular}%
\right) -\left(
\begin{tabular}{c}
$.5$ \\
$.5$%
\end{tabular}%
\right) =\left(
\begin{tabular}{ll}
$\phi $ & $0$ \\
$0$ & $\phi $%
\end{tabular}%
\right) \left( \left(
\begin{tabular}{l}
$X_{t-1}$ \\
$Y_{t-1}$%
\end{tabular}%
\right) -\left(
\begin{tabular}{c}
$.5$ \\
$.5$%
\end{tabular}%
\right) \right) +\left(
\begin{tabular}{l}
$\epsilon _{t}^{1}$ \\
$\epsilon _{t}^{2}$%
\end{tabular}%
\right) ,
\end{equation*}%
where $\left( \epsilon _{t}^{1},\epsilon _{t}^{2}\right) ^{\prime }$ are iid bivariate Gaussian distributed with zero mean and correlation parameter $\rho $. Three values of the correlation parameter $\rho $ are considered,
$\rho=-.5,$ $0$ and $.5$. Three values of the parameter $\phi $ are considered, $\phi =-.5$, $\phi =0$ and $\phi =.8$, to represent the case where $X_{t}$ and $Y_{t}$ are close to non-stationary. Sample sizes are $T=200$, $500$, $1000$, $2000$ and $3000$. Table~\ref{table1} gives the results based on $1000$ replications and an initial nominal significant level of $\alpha _{0}=0.05$. From this table, it seems that the type I error of the proposed procedure (the accuracy of the estimator of the number of change points if the true value is zero) is very close to the initial nominal level even with the smallest sample size. Therefore, overestimation does not appear to be an issue for the proposed procedure in this situation if there are no changes in the correlation.

\bigskip

\begin{center}
\textbf{{Table~\ref{table1} about here}}
\end{center}

\bigskip

Next, we analyze the power of our procedure when there is a single change point in the series. The Monte Carlo setup is similar to the one described above, but the series are generated with a single change point in the
correlation. Three locations of the change point are considered, $z_{1}=0.25$, $0.50$ and $0.75$. The change is such that $\rho$ is initially $\rho _{0}=.25$ and then changes to $\rho _{1}=-.25$, to represent a big change, to
$\rho _{1}=.15$, to represent a small change, or to $\rho _{1}=.5$, to represent a moderate change. Tables~\ref{table2}, \ref{table3} and \ref{table4} show the relative frequency detection of zero, one and more than one
changes. It is seen that the procedure performs quite well in detecting a single change point if the size of the change is moderate and large, with many cases over $90\%$ correct detection. However, if the size of the change is
small, then the power is small. Second, as the sample size increases and the size of the change gets larger, the procedure works better. However, the magnitudes of the exception are small in general. Third, when the sample size
of the change is small, the probability of under-detection may be large. Fourth, the location of the change point does not strongly affect the detection frequency of the procedure when the sample size is large. However, if the
sample size is small then the procedure detects more frequently the change point at the middle of the series. Finally, in most cases, the percentage of false detection is smaller than the nominal $5\%$. In particular, the
frequency of over-detection is small unless the two series are close to nonstationarity. On the other hand, Table~\ref{table5} shows the median and mean absolute deviation of the change point estimators in each case. The median of the estimates are quite close to the true change point locations. Note that the
larger the size of the change, the better is the location estimated.

\bigskip

\begin{center}
\textbf{{Table~\ref{table2} about here} }

\textbf{{Table~\ref{table3} about here} }

\textbf{{Table~\ref{table4} about here} }

\textbf{{Table~\ref{table5} about here} }
\end{center}

\bigskip

Next, we conduct another Monte Carlo experiment to study the power of the proposed procedure for detecting two change points. In this case, the location of the change points are $z_{1}=0.25$ and $z_{2}=0.75$, respectively. Three situations are considered. First, the changes are such that the correlation of the series before the first change point is $\rho _{0}=.25$, then changes to $\rho _{1}=0$, and, finally, changes to $\rho_{2}=.25$ at the second
change point. Second, the correlation of the series before the first change point is $\rho _{0}=.25$, then changes to $\rho _{1}=.5$, and, finally, changes to $\rho _{2}=.0$ at the second change point. Third, the correlation of
the series before the first change point is $\rho_{0}=.25$, then changes to $\rho _{1}=0$, and, finally, changes to $\rho _{2}=.25$ at the second change point. It is important to note that the first and the third of the
situations do not fulfill Assumption \ref{ass:MomentsFunctiong}. However, we consider these situations in order to show that, even if there is not a dominating change point, the procedure appears to perform well in these
situations. Indeed, the results suggest that the procedure consistently estimates the number of change points even if Assumption \ref{ass:MomentsFunctiong} does not hold. Table~\ref{table6} shows the relative frequency detection of zero, one, two and more than two changes. As in the case of a single change point, the proposed procedure works reasonably well, especially when the sample size is large or the size of the correlation change is large. In
addition, the procedure does not overestimate the number of change points. It may underestimate the number of change points, however. The underestimation can be serious when the sample size is small, say $T=200$, which indicates that the procedure has to be applied with care for small sample size. Finally, the percentage of false change points detected in both cases, one and two change points, is smaller than the nominal $5\%$ in almost all the cases. On the other hand, Table~\ref{table7} shows the median and mean absolute deviation of the estimates of the change point locations. Note that the medians of the estimates are quite close to the true ones. Again, it appears that the
larger is the size of the change, the better is the location estimated.

\bigskip

\begin{center}
\textbf{{Table~\ref{table6} about here} }

\textbf{{Table~\ref{table7} about here} }
\end{center}

\bigskip

To finish with the VAR(1) model, we repeat the previous experiment when there are two change points in the correlation of the series plus two change points in the mean of the series. The Monte Carlo setup is similar to the one
described above, but the mean of the series pass from $\left(.5,.5\right) ^{\prime }$ to $\left(1,1\right) ^{\prime }$. The results are shown in Tables~\ref{table8} and ~\ref{table9}. We deduce from these results that the
inclusion of changes in the mean additionally to changes in the correlation does not affect importantly the results of the procedure. Note however that the size of the fluctuation test is usually affected by mean changes: If the correlation is constant and if there are mean changes, the empirical size is typically higher than the nominal size. Detailed results are available upon request.

\bigskip

\begin{center}
\textbf{{Table~\ref{table8} about here} }

\textbf{{Table~\ref{table9} about here} }
\end{center}

\bigskip

For the second scenario, more appropriate for financial returns, initially we check the size (probability of a type I error) of the procedure which can be equivalently considered as the accuracy of the estimator of the number of change points if the true value is zero. To this purpose, we consider a dynamic conditional correlation model as in \citet{tsetsu:2002} given by:
\begin{equation*}
\left(
\begin{tabular}{l}
$X_{t}$ \\
$Y_{t}$%
\end{tabular}%
\right) =\left(
\begin{tabular}{cc}
$H_{X,t}$ & $H_{XY,t}$ \\
$H_{XY,t}$ & $H_{Y,t}$%
\end{tabular}%
\right) ^{1/2}\left(
\begin{tabular}{l}
$\epsilon _{t}^{1}$ \\
$\epsilon _{t}^{2}$%
\end{tabular}%
\right)
\end{equation*}%
where
\begin{equation*}
\left(
\begin{tabular}{cc}
$H_{X,t}$ & $H_{XY,t}$ \\
$H_{XY,t}$ & $H_{Y,t}$%
\end{tabular}%
\right) =\left(
\begin{tabular}{cc}
$H_{X,t}^{1/2}$ & $0$ \\
$0$ & $H_{Y,t}^{1/2}$%
\end{tabular}%
\right) \left(
\begin{tabular}{cc}
$1$ & $R_{XYt}$ \\
$R_{XYt}$ & $1$%
\end{tabular}%
\right) \left(
\begin{tabular}{cc}
$H_{X,t}^{1/2}$ & $0$ \\
$0$ & $H_{Y,t}^{1/2}$%
\end{tabular}%
\right) ,
\end{equation*}%
with $H_{X,t}=10^{-4}+0.1X_{t-1}+0.85H_{X,t-1}$ and $H_{Y,t}=10^{-4}+0.15Y_{t-1}+0.8H_{Y,t-1}$ are the individual volatilities of $X_{t}$ and $Y_{t}$ respectively, driven by univariate GARCH models, and:
\begin{equation*}
R_{XYt}=\left( 1-.95-.03\right) \rho+.95R_{XY,t-1}+.03\times U
\end{equation*}%
is the conditional correlation between $X_{t}$ and $Y_{t}$, $-1<\rho<1$, $U$ is the $2\times 2$ matrix of ones and $\left( \epsilon _{t}^{1},\epsilon _{t}^{2}\right)^{\prime }$ are iid bivariate standard Gaussian distributed. Three values of the correlation parameter $\rho $ are considered, $\rho=0$, $0.5$ and $0.8$, to represent the cases where $X_{t}$ and $Y_{t}$ are uncorrelated and have medium and high unconditional correlation, respectively.
Sample sizes are $T=500$, $1000$, $2000$, $3000$ and $4000$ that represent usual sample sizes in financial returns. Table~\ref{table10} shows the relative frequency detection of zero and more than zero changes based on $1000$ replications and an initial nominal significant level of $\alpha _{0}=0.05$. From this table, as in the VAR(1) case, it seems that the type I error of the proposed procedure is very close to the initial nominal level.

\bigskip

\begin{center}
\textbf{{Table~\ref{table10} about here}}
\end{center}

\bigskip

Next, we analyze the power of our procedure when there is a single change point in the series. The Monte Carlo setup is similar to the one described above, but the series are generated with a single change point in the
correlation. Three locations of the change point are considered, $z_{1}=0.25$, $0.50$ and $0.75$. The change is such that the correlation of the series is $\rho _{0}=.5$ and then changes to $\rho _{1}=.6$, $.7$ or $.8$, that
represent a small, moderate and high correlation change, respectively. Table~\ref{table11} shows the relative frequency detection of zero, one and more than one changes. The results appear to confirm the conclusions given in the case of the VAR(1) model. Consequently, note that conditional variances and correlations does not appear to affect the power of the procedure for detecting one change point in the correlation structure of the two series. In particular, note that even a small change can be reasonably well detected by the procedure. On the other hand, Table~\ref{table12} shows the median and mean absolute deviation of the change point estimators in each case and
appears to confirm that the larger the size of the change, the better is the location estimated.

\bigskip

\begin{center}
\textbf{{Table~\ref{table11} about here} }

\textbf{{Table~\ref{table12} about here} }
\end{center}

\bigskip

Finally, we conduct a Monte Carlo experiment to study the power of the proposed procedure for detecting two change points for the DCC model. As in the VAR(1) case, the location of the change points are $z_{1}=0.25$ and
$z_{2}=0.75$. Three situations are considered. First, the changes are such that the correlation of the series before the first change point is $\rho_{0}=.5$, then changes to $\rho _{1}=.7$, and, finally, changes to $\rho_{2}=.5$ at the second change point. Second, the correlation of the series before the first change point is $\rho _{0}=.5$, then changes to $\rho_{1}=.7$, and, finally, changes to $\rho _{2}=.6$ at the second change point. Third, the
correlation of the series before the first change point is $\rho _{0}=.5$, then changes to $\rho _{1}=.6$, and, finally, changes to $\rho _{2}=.7$  at the second change point. Again, it is important to note that the first
situation does not fulfill Assumption \ref{ass:MomentsFunctiong}, but we include it to show that even if there is not a dominating change point, the procedure appears to perform well. Table~\ref{table13} shows the relative
frequency detection of zero, one, two and more than two changes. As in the VAR(1) case, the proposed procedure works reasonably well, especially when the sample size is large or the size of the correlation change is large. In
particular, note that the procedure does not overestimate the number of change points and that it may underestimate the number of change points only when both the size of the changes and the sample size are small. On the other
hand, Table~\ref{table14} shows the median and mean absolute deviation of the estimates of the change point locations leading to similar conclusions as before.

\bigskip

\begin{center}
\textbf{{Table~\ref{table13} about here} }

\textbf{{Table~\ref{table14} about here} }
\end{center}

\bigskip

\section{Application}
\label{sec:appl}

\begin{figure}[!tbp]
\centering
\caption{S\&P 500 and IBM log-returns} \label{Figure3}
\includegraphics[width=6.0in,height=3.5in]{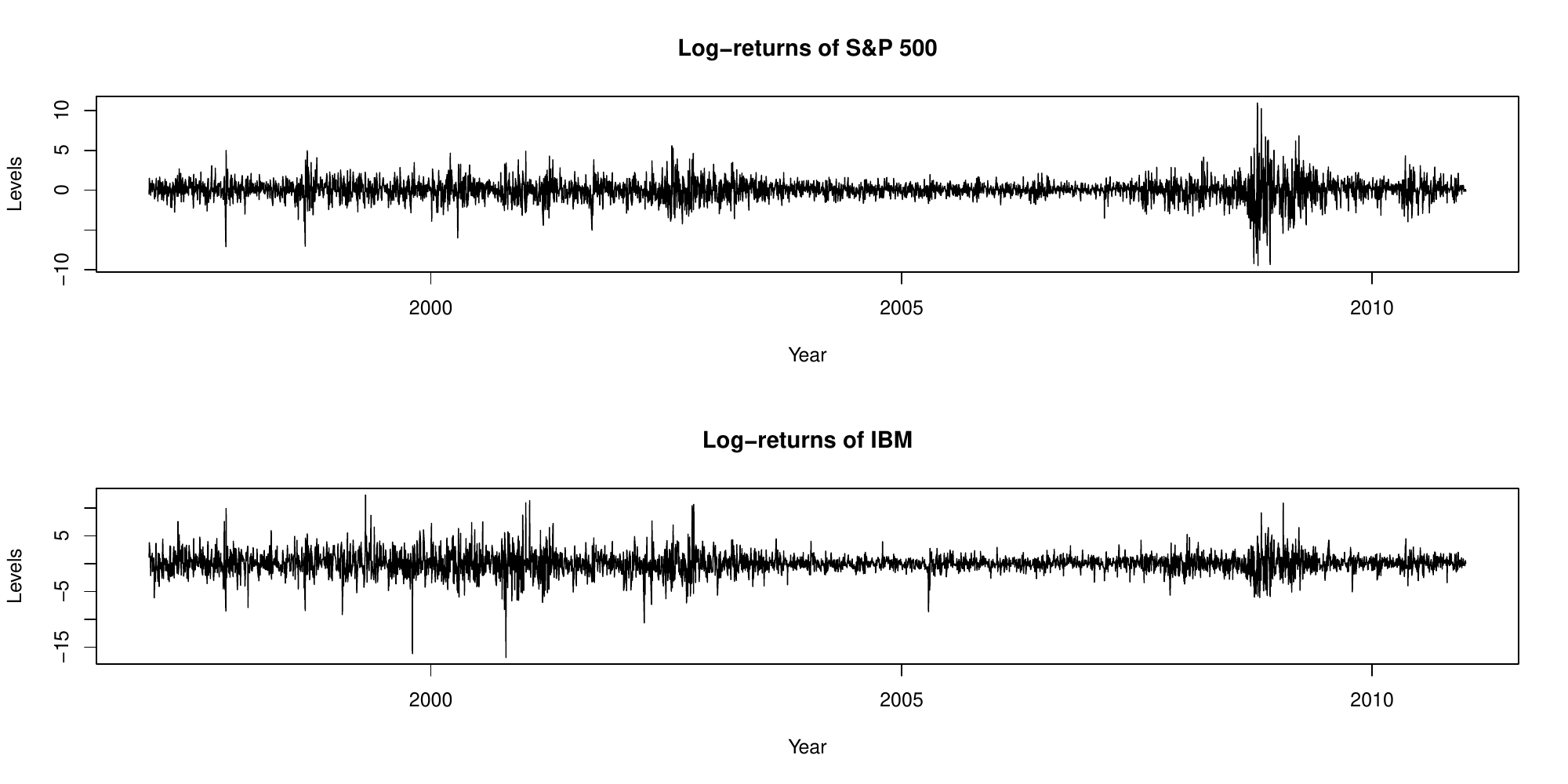}
\end{figure}

This section looks for changes in the correlation structure of the log-return series of the Standard \&\ Poors 500 Index and the IBM stock from January 2, 1997 to December 31, 2010 consisting of $T=3524$ data points. Both log-returns series are plotted in Figure~\ref{Figure3}, which shows
different volatility periods. The empirical full sample correlation is $0.6225$. The autocorrelation functions of the log-returns show some minor
serial dependence, while the autocorrelation functions of the squared log-returns reveals considerable serial dependence, as usual in stock market
returns.

\begin{center}
\bf{Figure \ref{Figure3} about here}
\end{center}

\begin{figure}[!tbp]
\centering
\caption{First step of the procedure} \label{Figure4}
\includegraphics[width=6.0in,height=3.5in]{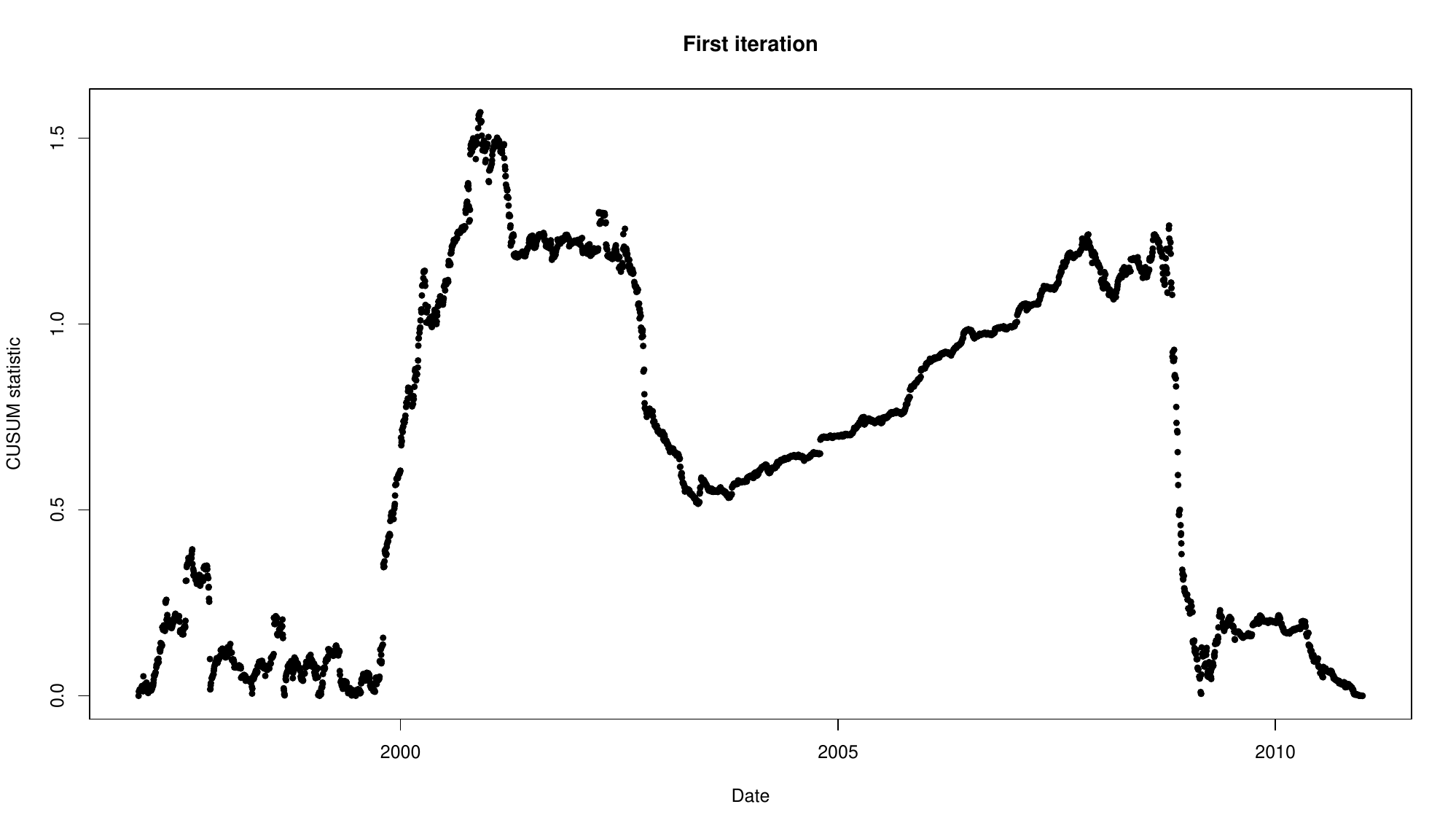}
\end{figure}

\begin{figure}[!tbp]
\centering
\caption{Second step of the procedure (first iteration)} \label{Figure5}
\includegraphics[width=6.0in,height=3.5in]{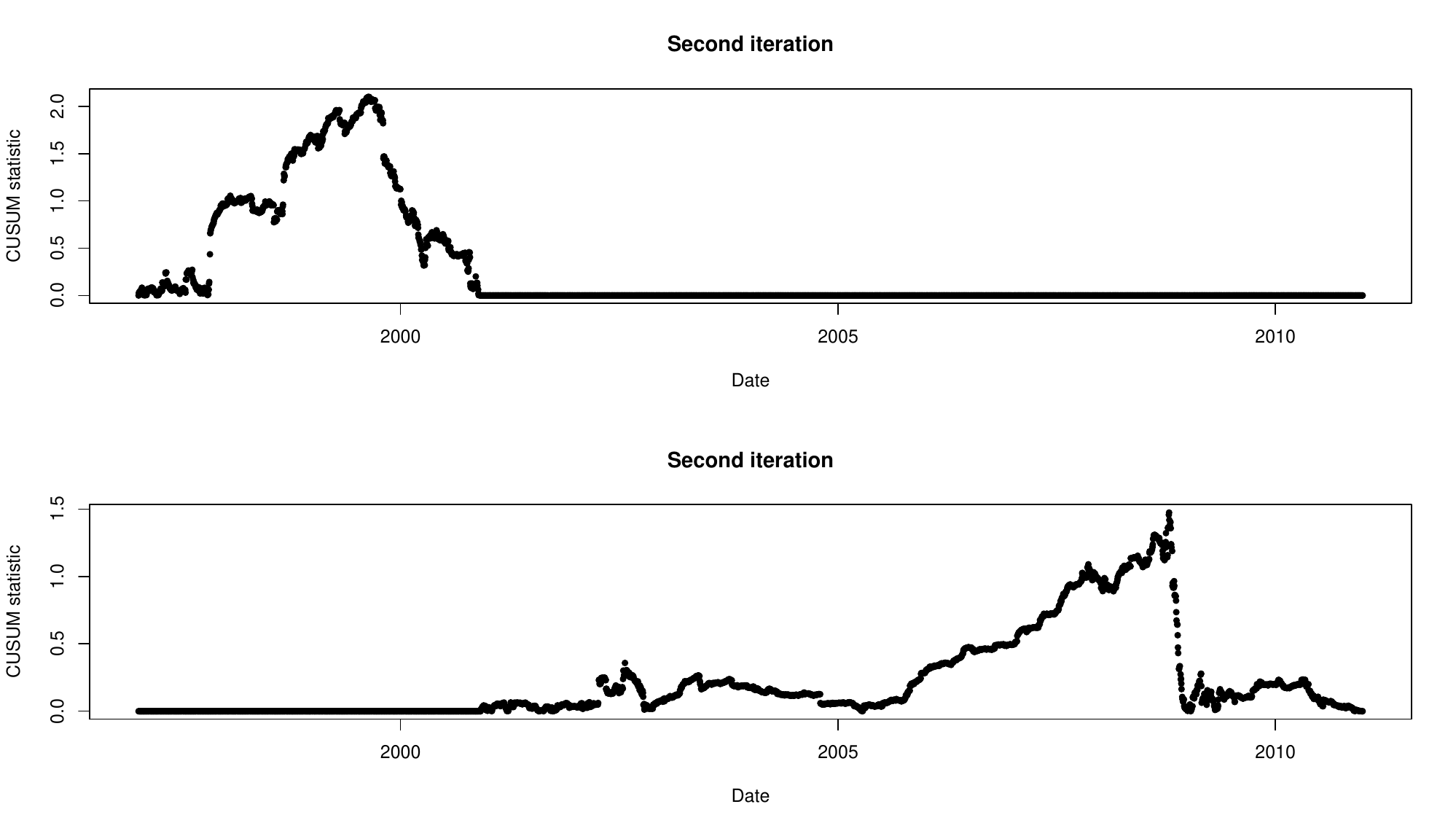}
\end{figure}

\begin{figure}[!tbp]
\centering
\caption{Second step of the procedure (second iteration)} \label{Figure6}
\includegraphics[width=6.0in,height=3.5in]{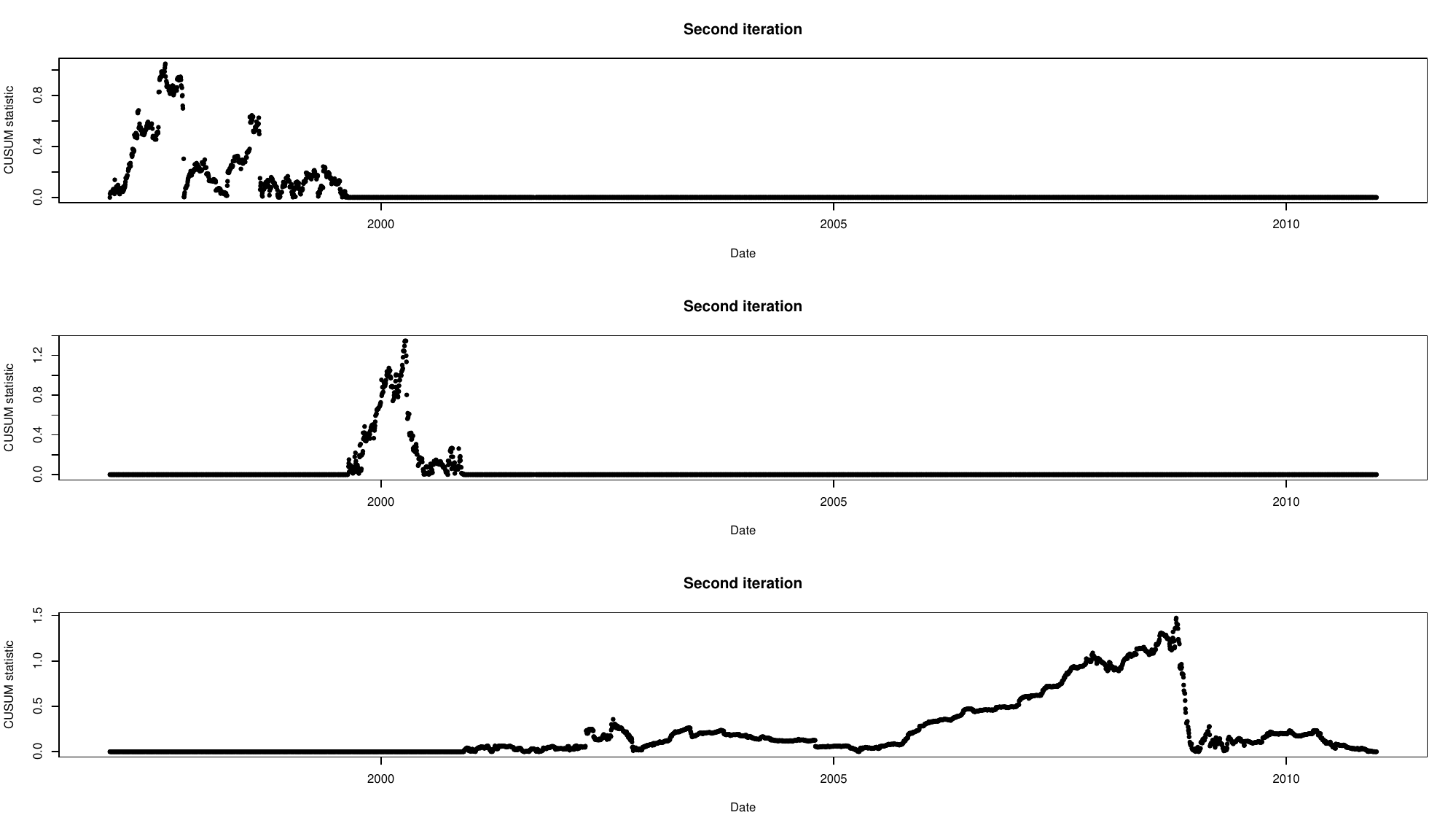}
\end{figure}

\begin{figure}[!tbp]
\centering
\caption{Third step of the algorithm} \label{Figure7}
\includegraphics[width=6.0in,height=3.5in]{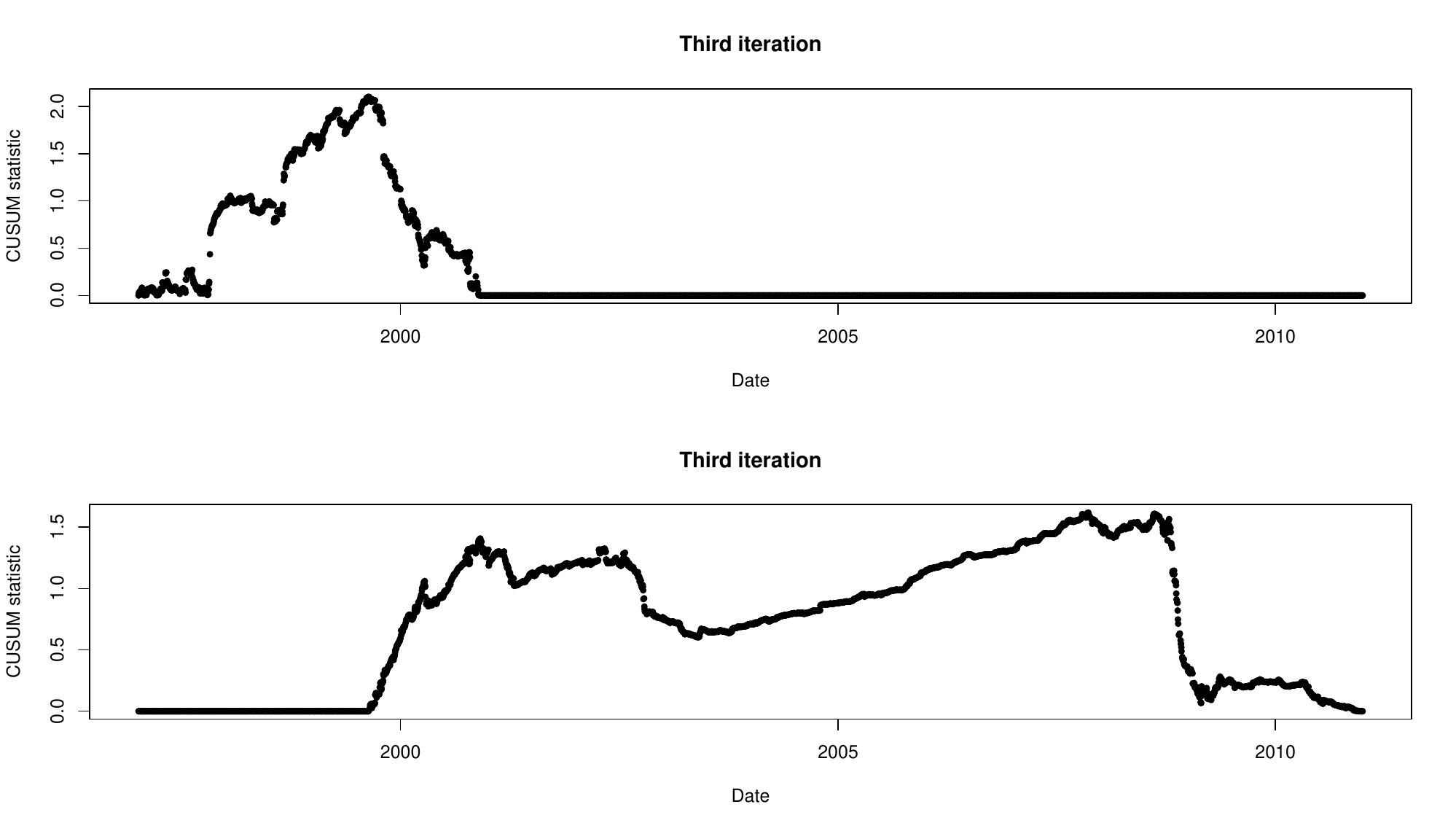}
\end{figure}

Next, we apply the proposed segmentation procedure of Section 2 to detect correlation changes for the log-returns of the S\&P 500 and IBM stock
assets. Table~\ref{table15} and Figures~\ref{Figure4}, ~\ref{Figure5}, ~\ref{Figure6} and ~\ref{Figure7} show the iterations taken by the procedure.
Similar to the simulation experiments of Section \ref{sec:finitesample}, we start with the asymptotic critical value at the 5\% significance level. In the first iteration, the procedure detects a change in the correlation at time point $t=988$ (November 29, 2000). Indeed, as shown in Figure~\ref{Figure4} there are two local modes of the CUSUM statistic. The value of the test statistic (\ref{Q_T}) is $1.5700$, which is statistically significant at the $5\%$ level. Following the proposed procedure, we split the series into two subperiods and look for changes in the subintervals $\left[1, 988\right] $ and $\left[989,3524\right]$, respectively. In the first subinterval (see Figure~\ref{Figure5}), the procedure detects a change at time point $t=664$ (August 19, 1999). The value of the test statistic is $2.1009$. Then, we split the subinterval $\left[1,988\right]$ into two subintervals and look for changes in the subintervals $\left[1, 664\right]$, $\left[665,988\right]$ and $\left[989,3524\right]$ (see Figure~\ref{Figure6}). No more changes were found in these three subintervals. Then, we pass to step 3 and refine the search. For that we estimate the location of the change points in the intervals $\left[1, 988\right] $ and $\left[665,3524\right]$, respectively. In the first subinterval (see Figure~\ref{Figure7}), as in the previous step, the procedure detects a change at time point $t=664$ (August 19, 1999) and the value of the test statistic is $2.1009$. On the other hand, in the second subinterval (see Figure~\ref{Figure7}), as in the previous step, the procedure detects a change at time point $t=2734$ (November 12, 2007) and the value of the test statistic is $1.6193$. These are the finally estimated change points.

\bigskip

\begin{center}
\bf{Table~\ref{table15} about here}

\bf{Figure~\ref{Figure4} about here}

\bf{Figure~\ref{Figure5} about here}

\bf{Figure~\ref{Figure6} about here}

\bf{Figure~\ref{Figure7} about here}
\end{center}

\bigskip

\begin{figure}[!tbp]
\centering
\caption{Scatterplots of the two log-returns at three different subperiods} \label{Figure8}
\includegraphics[scale=0.5]{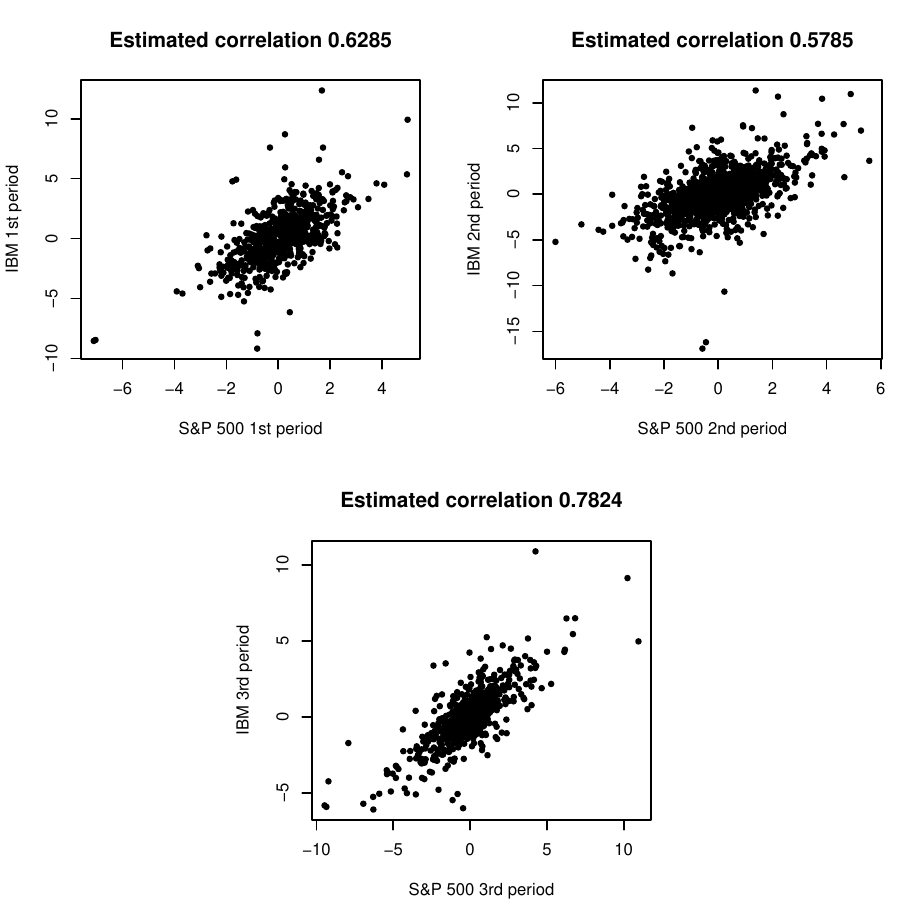}
\end{figure}

The empirical correlation coefficients in the three subintervals are $0.6285$, $0.5785$ and $0.7824$, respectively, indicating that the correlation shifted to a smaller value after the first change point and to a higher value
after the second change point. Figure~\ref{Figure8} shows the scatterplots of the two log-returns indexes at three different subperiods. It is interesting to see that the dates of the detected change points fare well with well
known financial facts. The period starting at $1994$ till the end of $1999$ is a period of economic growth in the U.S. economy in which the inflation was under control and the unemployment rate dropped to below $5\%$. This is a
period with high increases in the stock markets. However, the collapse of the dot-com bubble started at the end of the 1990s and the beginning of the 2000s, and the market gave back around the $75\%$ of the growth obtained in the 1990s. However, note that, contrarily to the diversification meltdown theory, the correlation did not increase during the dot-com bubble crisis. The third estimated change point roughly corresponds to the beginning of the Global Financial Crisis around the end of 2007, which is considered by many economists the worst financial crisis since the Great Depression of the 1930s. The reduction of interest rates leads to several consequent issues starting with the easiness of obtaining credit, leading to sub-prime lending, so that an increased debt burden, and finally a liquidity shortfall in the banking system. This resulted in the collapse of well known financial institutions such as Lehman Brothers, Merrill Lynch, Washington Mutual, Wachovia, and AIG, among others, the bailout of banks by national governments such as Bear Stearns, Citigroup, Bank of America and Northern Rock, among others, and great loses in stock markets around the world. In this case, the Global Financial Crisis produced an increase in the correlation between both log-returns. Of course, it is important to note that these economic interpretations are mere speculations. These comments only point out that, for this particular example, the proposed detection procedure in Section~\ref{sec:algorithm} identifies changes in the correlation structure that fare well with well known events affecting the U.S. financial market.

\bigskip

\begin{center}
\bf{Figure~\ref{Figure8} about here}
\end{center}

\bigskip

\section{Conclusions}
\label{sec:conc}

In this paper, we have proposed a binary segmentation procedure for change points in the correlation structure of random variables. As far as we know, this is the first procedure for solving such a problem. The procedure is based on a CUSUM test statistic proposed by \citet{wied:2011}. The asymptotic distribution of the test coincides with the one of the supremum of the absolute value of a standard Brownian bridge in the interval $[0,1]$. We have
shown that the proposed procedure consistently detects the true number and the location of the change points. Also, the finite sample properties of
the procedure have been analyzed by the analysis of several simulation studies and the application of the procedure to a real data example. The
empirical findings in the real data example suggest that the procedure detects changes in situations in which the relationship between financial
returns may change due to financial crisis. Some care is necessary if the procedure is applied on datasets with small sample size.

A potential drawback of our procedure is the fact that it is designed for a bivariate vector only. Of course it is possible to consider each entry in a higher dimensional correlation matrix separately to determine whether there have been changes in the individual correlations. It might be an
interesting issue for further research to do this in a more sophisticated way. Moreover, it might be interesting to consider other methods of
detecting multiple breaks in correlation, for example a method building on the simultaneous method in \citet{bai:1998}. This method would be quite
different as the break points would not be estimated step-by-step, but by performing one single minimization: Given a certain sum of squares and the
number $\ell$ of break points, one searches for the $\ell$ break points which minimize this sum. It would be interesting to see which procedure
provides advantages in which situation. \\

{\it Acknowledgements:} Financial support by MCI grants MTM2008-03010 and ECO2012-38442 and Deutsche Forschungsgemeinschaft (SFB 823, project A1) is gratefully acknowledged. We are grateful to the referees for helpful suggestions. The paper also benefitted a lot from detailed comments by Walter Krämer.

\bibliography{galeano_wied}

\appendix

\section{Appendix}
\subsection{The scalar $\hat D$ from the test statistic \eqref{Q_T}}
~\\
The scalar $\hat D$ from our test statistic $Q_T(X,Y)$ based on all observations can be written as
\begin{align*}
\hat D=(\hat{F}_1 \hat{D}_{3,1} + \hat{F}_2 \hat{D}_{3,2} + \hat{F}_3 \hat{D}_{3,3})^{-\frac{1}{2}}
\end{align*}
where
\begin{align*}
\begin{pmatrix} \hat{F}_1 & \hat{F}_2 & \hat{F}_3 \end{pmatrix} = \begin{pmatrix} \hat{D}_{3,1} \hat{E}_{11} + \hat{D}_{3,2} \hat{E}_{21} + \hat{D}_{3,3} \hat{E}_{31} \\ \hat{D}_{3,1} \hat{E}_{12} + \hat{D}_{3,2} \hat{E}_{22} + \hat{D}_{3,3} \hat{E}_{32} \\ \hat{D}_{3,1} \hat{E}_{13} + \hat{D}_{3,2} \hat{E}_{23} + \hat{D}_{3,3} \hat{E}_{33} \end{pmatrix}',
\end{align*}
\begin{align*}
\hat{E}_{11} &= \hat{D}_{1,11} - 4 \hat{\mu}_{x} \hat{D}_{1,13} + 4 {\hat{\mu}_{x}}^2 \hat{D}_{1,33}, \\
\hat{E}_{12} &= \hat{E}_{21} = \hat{D}_{1,12} - 2 \hat{\mu}_{x} \hat{D}_{1,23} - 2 \hat{\mu}_{y} \hat{D}_{1,14} + 4 \hat{\mu}_{x} \hat{\mu}_{y} \hat{D}_{1,34}, \\
\hat{E}_{22} &= \hat{D}_{1,22} - 4 \hat{\mu}_{y} \hat{D}_{1,24} + 4 {\hat{\mu}_{y}}^2 \hat{D}_{1,44}, \\
\hat{E}_{13} &= \hat{E}_{31} = - \hat{\mu}_{y} \hat{D}_{1,13} + 2 \hat{\mu}_{x} \hat{\mu}_{y} \hat{D}_{1,33} - \hat{\mu}_{x} \hat{D}_{1,14} + 2 {\hat{\mu}_{x}}^2 \hat{D}_{1,34} + \hat{D}_{1,15} - 2 \hat{\mu}_{x} \hat{D}_{1,35},  \\
\hat{E}_{23} &= \hat{E}_{32} = - \hat{\mu}_{y} \hat{D}_{1,23} + 2 \hat{\mu}_{x} \hat{\mu}_{y} \hat{D}_{1,44} - \hat{\mu}_{x} \hat{D}_{1,24} + 2 {\hat{\mu}_{y}}^2 \hat{D}_{1,34} + \hat{D}_{1,25} - 2 \hat{\mu}_{y} \hat{D}_{1,45},  \\
\hat{E}_{33} &= {\hat{\mu}_{y}}^2 \hat{D}_{1,33} + 2 \hat{\mu}_{x} \hat{\mu}_{y} \hat{D}_{1,34} - 2 \hat{\mu}_{y} \hat{D}_{1,35} + {\hat{\mu}_{x}}^2 \hat{D}_{1,44} + \hat{D}_{1,55} - 2 \hat{\mu}_{x} \hat{D}_{1,45},
\end{align*}
\begin{align*}
\hat{D}_1 = \begin{pmatrix} \hat{D}_{1,11} & \hat{D}_{1,12} & \hat{D}_{1,13} & \hat{D}_{1,14} & \hat{D}_{1,15} \\
                           \hat{D}_{1,21} & \hat{D}_{1,22} & \hat{D}_{1,23} & \hat{D}_{1,24} & \hat{D}_{1,25} \\
                           \hat{D}_{1,31} & \hat{D}_{1,32} & \hat{D}_{1,33} & \hat{D}_{1,34} & \hat{D}_{1,35} \\
                           \hat{D}_{1,41} & \hat{D}_{1,42} & \hat{D}_{1,43} & \hat{D}_{1,44} & \hat{D}_{1,45} \\
                           \hat{D}_{1,51} & \hat{D}_{1,52} & \hat{D}_{1,53} & \hat{D}_{1,54} & \hat{D}_{1,55} \end{pmatrix}
                           = \sum_{t=1}^T \sum_{u=1}^T k\left(\frac{t-u}{\gamma_T}\right) V_t {V_u}',
\end{align*}
\begin{align*}
V_t = \frac{1}{\sqrt{T}} U^{***}_t, \gamma_T = [\log T],
\end{align*}
\begin{align*}
U^{***}_t = \begin{pmatrix} X_t^2 - \overline{(X^2)}_T & Y_t^2 - \overline{(Y^2)}_T & X_t - \bar X_T & Y_t - \bar Y_T & X_t Y_t - \overline{(XY)}_T \end{pmatrix}',
\end{align*}
\begin{align*}
k(x)=\begin{cases} 1 - |x|, & |x| \leq 1 \\ 0, & otherwise \end{cases},
\end{align*}
\begin{align*}
\hat{\mu}_{x} = \bar X_T,\hat{\mu}_{y} = \bar Y_T,\hat{D}_{3,1} &= -\frac{1}{2} \frac{\hat{\sigma}_{xy}}{\hat{\sigma}_y} \hat{\sigma}_x^{-3},\hat{D}_{3,2} = -\frac{1}{2} \frac{\hat{\sigma}_{xy}}{\hat{\sigma}_x} \hat{\sigma}_y^{-3},\hat{D}_{3,3} = \frac{1}{\hat{\sigma}_x \hat{\sigma}_y}
\end{align*}
and
\begin{align*}
{\hat{\sigma}_x}^2 &= \overline{(X^2)}_T - (\bar X_T)^2,{\hat{\sigma}_y}^2 = \overline{(Y^2)}_T - (\bar Y_T)^2,\hat{\sigma}_{xy} = \overline{(XY)}_T - \bar X_T \bar Y_T.
\end{align*}
This is the same expression as in Appendix A.1 in \citet{wied:2011}. The estimator based on the relevant sub-sample is basically calculated in the same way with the respective interval length taken into account. \\
\subsection{Proof of theorems}
~\\
{\it Proof of Theorem \ref{theorem:ConsistencySeveralBreaks}} \\
We consider the interval $[l_1,l_2]$ and consider $z \in [l_1,l_2]$. Denote
\begin{equation*}
\begin{split}
&\overline{X}_{\eta(l_1)}^{\xi(z)} = \frac{1}{\xi(z)-\eta(l_1)+1} \sum_{t=\eta(l_1)}^{\xi(z)} X_t,\
\overline{Y}_{\eta(l_1)}^{\xi(z)} = \frac{1}{\xi(z)-\eta(l_1)+1} \sum_{t=\eta(l_1)}^{\xi(z)} Y_t, \\
&\overline{XY}_{\eta(l_1)}^{\xi(z)} = \frac{1}{\xi(z)-\eta(l_1)+1} \sum_{t=\eta(l_1)}^{\xi(z)} X_t Y_t, \\
&\overline{[\mathsf{Var} X]}_{\eta(l_1)}^{\xi(z)} = \frac{1}{\xi(z)-\eta(l_1)+1} \sum_{t=\eta(l_1)}^{\xi(z)} \left(X_t - \overline{X}_{\eta(l_1)}^{\xi(z)}\right)^2,\\
&\overline{[\mathsf{Var} Y]}_{\eta(l_1)}^{\xi(z)} = \frac{1}{\xi(z)-\eta(l_1)+1} \sum_{t=\eta(l_1)}^{\xi(z)} \left(Y_t - \overline{Y}_{\eta(l_1)}^{\xi(z)}\right)^2,\\
&\hat \rho_{\eta(l_1)}^{\xi(z)} = \frac{\overline{XY}_{\eta(l_1)}^{\xi(z)} - \overline{X}_{\eta(l_1)}^{\xi(z)} \overline{Y}_{\eta(l_1)}^{\xi(z)}}{\sqrt{ [\mathsf{Var} X]_{\eta(l_1)}^{\xi(z)} } \sqrt{ [\mathsf{Var} Y]_{\eta(l_1)}^{\xi(z)} }}
\end{split}
\end{equation*}
Let $\hat D$ be the variance estimator from Appendix A.1 calculated from the observations $\eta(l_1)$ to $\xi(l_2)$.
It suffices to show consistency of $\hat z^* := \mathsf{argmax}_{l_1 \leq z \leq l_2} |A_T(z)|$ with
\begin{equation*}
A_{T}(z):=\hat D \frac{\xi(z) - \eta(l_1)+1}{\xi(l_2) - \eta(l_1)+1} \left( \hat \rho_{\eta(l_1)}^{\xi(z)} - \hat \rho_{\eta(l_1)}^{\xi(l_2)} \right),
\end{equation*}
because the difference between $\hat z^*$ and $\hat z = \xi(\hat z^*)/T$ is $O_p(1/T)$.
We first show that $A_T(z)$ converges in distribution to
\begin{equation*}
A(z) := C_A \left( \int_{l_1}^z g(t) dt - \frac{z - l_1}{l_2 - l_1} \int_{l_1}^{l_2} g(t) dt \right)
\end{equation*}
uniformly in $z \in [l_1,l_2]$ with a constant $C_A := \frac{D_1^A}{\sigma_x \sigma_y (l_2 - l_1)}$ (where $D_1^A$ is the limit from $\hat D$ under the alternative from Assumption \ref{ass:dhatAlternative}). For this purpose, write
\begin{equation*}
A_{T}(z)=\hat{D}\frac{\xi(z) - \eta(l_1)+1}{\xi(l_2) - \eta(l_1)+1}\left( \hat \rho_{\eta(l_1)}^{\xi(z)} - \rho_0\right) - \hat{D}\frac{\xi(z) - \eta(l_1)+1}{\xi(l_2) - \eta(l_1)+1}\left( \hat \rho_{\eta(l_1)}^{\xi(l_2)} - \rho_0\right)
\end{equation*}
with $\rho_0 = \frac{m_{xy} - \mu_x \mu_y}{\sigma_x \sigma_y}$ and consider an arbitrary $\epsilon > 0$ such that $l_1 + \epsilon < l_2$. We thus have
\begin{equation*}
\begin{split}
A_{T}(z) &= \hat{D}\frac{\xi(z) - \eta(l_1)+1}{\xi(l_2) - \eta(l_1)+1}\left( \frac{\overline{XY}_{\eta(l_1)}^{\xi(z)} - \overline{X}_{\eta(l_1)}^{\xi(z)} \overline{Y}_{\eta(l_1)}^{\xi(z)} }{\sqrt{\overline{[\mathsf{Var} X]}_{\eta(l_1)}^{\xi(z)} \overline{[\mathsf{Var} Y]}_{\eta(l_1)}^{\xi(z)}}} - \frac{m_{xy} - \mu_x \mu_y}{\sigma_x \sigma_y}\right) - \\
&\hat{D}\frac{\xi(z) - \eta(l_1)+1}{\xi(l_2) - \eta(l_1)+1}\left( \frac{\overline{XY}_{\eta(l_1)}^{\xi(l_2)} - \overline{X}_{\eta(l_1)}^{\xi(l_2)} \overline{Y}_{\eta(l_1)}^{\xi(l_2)} }{\sqrt{\overline{[\mathsf{Var} X]}_{\eta(l_1)}^{\xi(l_2)} \overline{[\mathsf{Var} Y]}_{\eta(l_1)}^{\xi(l_2)}}} - \frac{m_{xy} - \mu_x \mu_y}{\sigma_x \sigma_y}\right)
\end{split}
\end{equation*}
Straightforward calculations using the strong law of large numbers (Theorem 20.21 in \citealp{davidson:1994}), Slutzky's theorem, the fact that
\begin{equation*}
\begin{split}
&\sup_{z \in [l_1+\epsilon,l_2]} \left| \frac{1}{\xi(l_2) - \eta(l_1) + 1} \sum_{t=\eta(l_1)}^{\xi(z)} (\mathsf{E}(X_t Y_t) - m_{xy}) - \frac{1}{l_2 - l_1} \int_{l_1}^z g(t) dt \right| \\
= &\sup_{z \in [l_1+\epsilon,l_2]} \left| \frac{1}{\xi(l_2) - \eta(l_1) + 1} \sum_{t=\eta(l_1)}^{\xi(z)} g\left(\frac{t}{T}\right) - \frac{1}{l_2 - l_1} \int_{l_1}^z g(t) dt \right| \rightarrow 0
\end{split}
\end{equation*}
and the fact that $\sup_{z \in [l_1+\epsilon,l_2]}(\xi(z) - \eta(l_1) + 1) \rightarrow \infty$ yield
\begin{equation*}
A_T(z) \rightarrow_{a.s.} A(z)
\end{equation*}
and
\begin{equation*}
|A_T(z)| \rightarrow_{a.s.} |A(z)|
\end{equation*}
uniformly on $[l_1+\epsilon,l_2]$.

%Consider now the case $\epsilon = 0$. From the preceding calculations, we immediately get
%\begin{equation*}
%A_T(z) \rightarrow_{a.s.} A(z)
%\end{equation*}
%uniformly on $[l_1+\epsilon,l_2]$ for a fixed $\epsilon > 0$ such that $l_1 + \epsilon < l_2$.

Consider now the following functions:
\begin{equation*}
A_T^{\epsilon}(z)=\begin{cases} A_T(z), & z \geq l_1 + \epsilon \\ 0 & l_1 \leq z < l_1 + \epsilon \end{cases},
\end{equation*}
\begin{equation*}
A^{\epsilon}(z)=\begin{cases} A(z), & z \geq l_1 + \epsilon \\ 0 & l_1 \leq z < l_1 + \epsilon \end{cases}.
\end{equation*}
As uniform almost sure convergence implies convergence in distribution, the previous results then imply that
\begin{equation*}
A_T^{\epsilon}(\cdot) \rightarrow_d A^{\epsilon}(\cdot)
\end{equation*}
for $T \rightarrow \infty$ on $[l_1+\epsilon,l_2]$ and also
\begin{equation*}
A^{\epsilon}(\cdot) \rightarrow_d A(\cdot)
\end{equation*}
for rational $\epsilon \rightarrow 0$. The convergence of $A_T(\cdot)$ on $[l_1,l_2]$ then follows from Theorem 4.2 in \cite{billingsley:1968} if we can show that
\begin{equation*}
\lim_{\epsilon \rightarrow 0} \limsup_{T \rightarrow \infty} \mathp(\sup_{z \in [l_1,l_2]}|A_T^{\epsilon}(z)-A_T(z)| \geq \eta) = \lim_{\epsilon \rightarrow 0} \limsup_{T \rightarrow \infty} \mathp(\sup_{z \in [l_1,l_1+\epsilon]}|A_T(z)| \geq \eta)=0
\end{equation*}
for all $\eta > 0$. Note that the separability condition of this theorem is not necessary in our case, because for each interval $I \subset [0,1]$, $\sup_{z \in I} |A(z)|$ is always a random variable when $A(\cdot)$ is a right-continuous random function.

Since
\begin{equation*}
\begin{split}
\limsup_{T \rightarrow \infty} \mathp(\sup_{z \in [l_1,l_1+\epsilon]}|A_T(z)| \geq \eta) &\leq \limsup_{T \rightarrow \infty} \mathp(\sup_{z \in [l_1,l_1+\epsilon]}\left|\hat D 2 \frac{\xi(z) - \eta(l_1)}{\xi(l_2) - \eta(l_1)+1}\right| \geq \eta) \\ &= \mathp(D_A 2 \epsilon \geq \eta)
\end{split}
\end{equation*}
with $D_A := \frac{D_1^A}{l_2-l_1}$ (where $D_1^A$ is from Assumption \ref{ass:dhatAlternative})
we get
\begin{equation*}
A_T(z) \rightarrow_{a.s.} A(z)
\end{equation*}
and
\begin{equation*}
|A_T(z)| \rightarrow_{a.s.} |A(z)|
\end{equation*}
uniformly on $[l_1,l_2]$.

With Assumptions \ref{ass:dominatingChangePoint} and \ref{ass:dhatAlternative}, $|A(z)|$ has a unique maximum $m$ in the change point fraction.
%(I think, we cannot prove anything here, we have to make that assumption)
Let $\hat F$ the maximum of $|A_T(z)|$ for $z \in [l_1,l_2]$ (and $z$ as multiples of $1/T$). Since $|A_T(\hat F)| \geq |A_T(m)|$ we get stochastic convergence of $\hat F$ to $m$ (compare the argument in \citealp{bai:1998}, p.77; also an application of the argmax continuous mapping theorem would be possible here). \hfill $\blacksquare$ \\

\noindent {\it Proof of Theorem \ref{theorem:AlgorithmWorks}} \\
Theorem 1 implies
\begin{equation*}
A_T(z) \rightarrow_{a.s.} A(z)
\end{equation*}
uniformly for $z \in [l_1,l_2]$ if there is a change point in the interval $[l_1,l_2]$. Denote $Q^k_T(X,Y)$ the test statistic calculated from data from $\eta(l_1)$ to $\xi(l_2)$. Since
\begin{equation*}
Q^k_T(X,Y) = \sqrt{\left( \xi(l_2) - \eta(l_1)+1 \right)} \sup_{z \in [l_1,l_2]} |A_T(z)|,
\end{equation*}
we have
\begin{equation}\label{Proof2}
\frac{1}{a^k_T} Q_T(X,Y) \rightarrow_p \infty
\end{equation}
for any sequence $a^k_T = o\left(\sqrt{T}\right)$ if there is a change point in the interval $[l_1,l_2]$.
With this argument (which is partially similar to Corollary 2 in \citealp{andrews:1993}), one can adapt the proof of Proposition 11 of \citet{bai:1997}.

Consider the event $\{\hat \ell < \ell\}$. If the estimated number of change points $\hat \ell$ is smaller than $\ell$, there is at least one segment $[\hat z_m, \hat z_n]$ with $\hat z_m \rightarrow_p z_m$ and $\hat z_n \rightarrow_p z_n$
such that there is another change point $z_o \in [z_m,z_n]$. Denote $Q^m_T(X,Y)$ the test statistic calculated from data from $\eta(\hat z_m)$ to $\xi(\hat z_n)$. Since $\mathsf{P}(Q^m_T(X,Y) > a^m_T) \rightarrow 1$ as $T \rightarrow \infty$ with \eqref{Proof2}, we have $\mathsf{P}(\hat \ell < \ell) \rightarrow 0$
as $T \rightarrow \infty$.
Consider the event $\{\hat \ell > \ell\}$. For this event to be true, there must be a false rejection of the null hypothesis at a certain stage in the segmentation procedure. If $(z_k, k = 0,\ldots,\ell)$ are the true
change points and $(\hat z_k, k = 0,\ldots,\ell)$ are the corresponding consistent estimates, it holds
\begin{equation*}
\begin{split}
\mathsf{P}(\hat \ell > \ell) &\leq \mathsf{P}(\exists k: \text{the test based on data for } \xi(z) \text{ with } z \in [\hat z_k,\hat z_{k+1}] \text{ rejects}) \\
                       &\leq \sum_{k=0}^{\ell} \mathsf{P}(\text{the test based on data for } \xi(z) \text{ with } z \in [\hat z_k,\hat z_{k+1}] \text{ rejects}).
\end{split}
\end{equation*}
Let $Q^i_T(X,Y)$ be the test statistic computed from data from $\eta(\hat z_i)$ to $\xi(\hat z_{i+1})$. Since under the null hypothesis $\mathsf{P}(Q^i_T(X,Y) > a^i_T) \rightarrow 0$, it holds
\begin{equation*}
\mathsf{P}(\hat \ell > \ell) \leq (\ell+1) \max_{0 \leq k \leq \ell} \mathsf{P}(Q^i_T(X,Y) > a^i_T) \rightarrow 0.
\end{equation*}
Consequently, $\mathsf{P}(\hat \ell \leq \ell) \rightarrow 1$ for $T \rightarrow \infty$.

Combining the argumentation for the event $\{\hat \ell < \ell\}$ with Theorem \ref{theorem:ConsistencySeveralBreaks} yields the proposed consistency results and the proof is completed.\hfill $\blacksquare$ \\

\noindent {\it Proof of Theorem \ref{theorem:AsyDis2}} \\

As in the proof of Theorem \ref{theorem:ConsistencySeveralBreaks}, we derive the limit of $\hat z^*$.
We make use of the fact that the quantitiy $\mathsf{argmax}_{l_1 \leq z \leq l_2} A_{T}(z)$ can be equivalently written as $\mathsf{argmax}_{l_1 \leq z \leq l_2} \sqrt{(l_2 - l_1) T} A_{T}(z)$.
This is important because we want to apply an argmax continuous mapping theorem later on which requires that the paths of the limit process almost surely have unique maxima.

In the first step, we adapt the proof of Theorem 2 in \citet{wied:2011} for the case that we consider the interval $[l_1,l_2]$ instead of the interval $[0,1]$. The basic difference is that we do not consider the convergence of the process
\begin{equation*}
\frac{1}{\sqrt{T}} \sum_{t=1}^{[zT]} (U_t - \mathsf{E}(U_t))
\end{equation*}
with $U_t$ from Assumption \ref{boundedmoments2} to a scaled Brownian motion $W(z)$ for $0 \leq z \leq 1$, but the convergence of the process
\begin{equation*}
\frac{1}{\sqrt{(l_2-l_1)T}} \sum_{t={[l_1 T]}}^{[zT]} (U_t - \mathsf{E}(U_t))
\end{equation*}
to $W(z) - W(l_1)$ for $l_1 \leq z \leq l_2$, which follows by the same functional central limit theorem as used in \citet{wied:2011}. With this adaption, we transfer the proof of Theorem 2 in \citet{wied:2011} and get convergence of the process $A_T(z) \sqrt{ \xi(l_2) - \eta(l_1)+1 }$ (with $A_T(z)$ from the proof of Theorem \ref{theorem:ConsistencySeveralBreaks}) to the process
\begin{equation}\label{limitprocessTheoremLocalPower}
W(z) - W(l_1) - \frac{z - l_1}{l_2 - l_1} \left( W(l_2) - W(l_1) \right) + D_A A^*(z)
\end{equation}
for $l_1 \leq z \leq l_2$. Here, $D_A := \frac{D_1^N}{l_2-l_1}$, where $D_1^N$ is the limit of $\hat D$ under the sequence of local alternatives that is equal to the limit of $\hat D$ under the null hypothesis to which the local alternatives converge.

The result of the theorem then follows with the argmax continuous mapping theorem from \citet{kimpollard:1990}, Theorem 2.7. This theorem can be applied here because it follows with Lemma 2.6 in \citet{kimpollard:1990}
that with probability $1$, every path of the Gaussian process \ref{limitprocessTheoremLocalPower} has a unique maximum. \hfill $\blacksquare$

\newpage

\begin{table}[tbp]
\caption{Type I errors with the VAR(1) model with a initial nominal significant level of $\alpha_0=0.05$.}
\begin{center}
$%
\begin{tabular}{cccccccc}
\cline{3-8}
&  & \multicolumn{2}{c}{$\rho _{0}=-.5$} & \multicolumn{2}{c}{$\rho _{0}=0$}
& \multicolumn{2}{c}{$\rho _{0}=.5$} \\ \cline{3-8}
&  & \multicolumn{2}{c}{Rel. freq.} & \multicolumn{2}{c}{Rel. freq.} &
\multicolumn{2}{c}{Rel. freq.} \\ \cline{3-8}
$\phi $ & $T$ & $0$ & $\geq 1$ & $0$ & $\geq 1$ & $0$ & $\geq 1$ \\ \hline
& $200$ & .949 & .051 & .952 & .048 & .925 & .075 \\
& $500$ & .949 & .051 & .947 & .053 & .954 & .046 \\
$-.5$ & $1000$ & .945 & .055 & .943 & .057 & .945 & .055 \\
& $2000$ & .953 & .047 & .955 & .045 & .946 & .054 \\
& $3000$ & .948 & .052 & .953 & .047 & .950 & .050 \\ \hline
& $200$ & .932 & .068 & .969 & .031 & .939 & .061 \\
& $500$ & .958 & .042 & .963 & .037 & .958 & .042 \\
$0$ & $1000$ & .955 & .045 & .958 & .042 & .970 & .030 \\
& $2000$ & .962 & .038 & .957 & .043 & .953 & .047 \\
& $3000$ & .954 & .046 & .956 & .044 & .962 & .038 \\ \hline
& $200$ & .786 & .214 & .798 & .202 & .791 & .209 \\
& $500$ & .830 & .170 & .841 & .159 & .832 & .168 \\
$0.8$ & $1000$ & .839 & .161 & .851 & .149 & .867 & .133 \\
& $2000$ & .886 & .114 & .857 & .143 & .859 & .141 \\
& $3000$ & .885 & .115 & .884 & .116 & .900 & .100 \\ \hline
\end{tabular}%
$%
\end{center}
\label{table1}
\end{table}

\begin{table}[tbp]
\caption{Relative frequency detection of 0, 1 and more than 1 change points with the VAR(1) model with a single change point for $z_1=.25$
and with a initial nominal significant level of $\alpha_0=0.05$.}
\begin{center}
$%
\begin{tabular}{ccccccccccc}
\cline{3-11}
&  & \multicolumn{3}{c}{$\left( \rho _{0}=.25,\rho _{1}=-.25\right) $} &
\multicolumn{3}{c}{$\left( \rho _{0}=.25,\rho _{1}=.15\right) $} &
\multicolumn{3}{c}{$\left( \rho _{0}=.25,\rho _{1}=.5\right) $} \\
\cline{3-11}
&  & \multicolumn{3}{c}{Rel. freq.} & \multicolumn{3}{c}{Rel. freq.} &
\multicolumn{3}{c}{Rel. freq.} \\ \cline{3-11}
$\phi $ & $T$ & $0$ & $1$ & $\geq 2$ & $0$ & $1$ & $\geq 2$ & $0$ & $1$ & $%
\geq 2$ \\ \hline
& $200$ & $.594$ & $.397$ & $.009$ & $.953$ & $.044$ & $.003$ & $.783$ & $%
.214$ & $.003$ \\
& $500$ & $.122$ & $.850$ & $.028$ & $.919$ & $.081$ & $.000$ & $.567$ & $%
.431$ & $.002$ \\
$-.5$ & $1000$ & $.001$ & $.941$ & $.058$ & $.881$ & $.109$ & $.010$ & $.244$
& $.745$ & $.011$ \\
& $2000$ & $.000$ & $.948$ & $.052$ & $.781$ & $.212$ & $.007$ & $.025$ & $%
.947$ & $.028$ \\
& $3000$ & $.000$ & $.924$ & $.076$ & $.718$ & $.268$ & $.014$ & $.003$ & $%
.938$ & $.059$ \\ \hline
& $200$ & $.409$ & $.584$ & $.007$ & $.950$ & $.050$ & $.000$ & $.707$ & $%
.291$ & $.002$ \\
& $500$ & $.020$ & $.957$ & $.023$ & $.924$ & $.074$ & $.002$ & $.369$ & $%
.623$ & $.008$ \\
$0$ & $1000$ & $.000$ & $.969$ & $.031$ & $.851$ & $.147$ & $.002$ & $.090$
& $.894$ & $.016$ \\
& $2000$ & $.000$ & $.939$ & $.061$ & $.681$ & $.309$ & $.010$ & $.002$ & $%
.977$ & $.021$ \\
& $3000$ & $.000$ & $.944$ & $.056$ & $.539$ & $.444$ & $.017$ & $.000$ & $%
.958$ & $.042$ \\ \hline
& $200$ & $.593$ & $.317$ & $.090$ & $.783$ & $.171$ & $.046$ & $.657$ & $%
.279$ & $.064$ \\
& $500$ & $.349$ & $.577$ & $.074$ & $.842$ & $.137$ & $.021$ & $.617$ & $%
.341$ & $.042$ \\
$.8$ & $1000$ & $.113$ & $.800$ & $.087$ & $.818$ & $.156$ & $.026$ & $.434$
& $.524$ & $.042$ \\
& $2000$ & $.009$ & $.866$ & $.125$ & $.840$ & $.142$ & $.018$ & $.243$ & $%
.696$ & $.061$ \\
& $3000$ & $.001$ & $.837$ & $.162$ & $.751$ & $.221$ & $.028$ & $.137$ & $%
.787$ & $.076$ \\ \hline
\end{tabular}%
$%
\end{center}
\label{table2}
\end{table}

\begin{table}[tbp]
\caption{Relative frequency detection of 0, 1 and more than 1 change points with the VAR(1) model with a single change point for $z_1=.5$
and with a initial nominal significant level of $\alpha_0=0.05$.}
\begin{center}
$%
\begin{tabular}{ccccccccccc}
\cline{3-11}
&  & \multicolumn{3}{c}{$\left( \rho _{0}=.25,\rho _{1}=-.25\right) $} &
\multicolumn{3}{c}{$\left( \rho _{0}=.25,\rho _{1}=.15\right) $} &
\multicolumn{3}{c}{$\left( \rho _{0}=.25,\rho _{1}=.5\right) $} \\
\cline{3-11}
&  & \multicolumn{3}{c}{Rel. freq.} & \multicolumn{3}{c}{Rel. freq.} &
\multicolumn{3}{c}{Rel. freq.} \\ \cline{3-11}
$\phi $ & $T$ & $0$ & $1$ & $\geq 2$ & $0$ & $1$ & $\geq 2$ & $0$ & $1$ & $%
\geq 2$ \\ \hline
& $200$ & $.364$ & $.617$ & $.019$ & $.931$ & $.067$ & $.002$ & $.969$ & $%
.295$ & $.009$ \\
& $500$ & $.018$ & $.941$ & $.041$ & $.887$ & $.108$ & $.005$ & $.379$ & $%
.609$ & $.012$ \\
$-.5$ & $1000$ & $.000$ & $.959$ & $.041$ & $.775$ & $.214$ & $.011$ & $.107$
& $.867$ & $.026$ \\
& $2000$ & $.000$ & $.936$ & $.064$ & $.614$ & $.360$ & $.026$ & $.005$ & $%
.952$ & $.043$ \\
& $3000$ & $.000$ & $.946$ & $.054$ & $.490$ & $.488$ & $.022$ & $.000$ & $%
.957$ & $.043$ \\ \hline
& $200$ & $.154$ & $.823$ & $.023$ & $.952$ & $.047$ & $.001$ & $.570$ & $%
.426$ & $.004$ \\
& $500$ & $.001$ & $.975$ & $.024$ & $.846$ & $.149$ & $.005$ & $.201$ & $%
.793$ & $.006$ \\
$0$ & $1000$ & $.000$ & $.963$ & $.037$ & $.710$ & $.279$ & $.011$ & $.019$
& $.962$ & $.019$ \\
& $2000$ & $.000$ & $.949$ & $.051$ & $.472$ & $.509$ & $.019$ & $.000$ & $%
.971$ & $.029$ \\
& $3000$ & $.000$ & $.941$ & $.059$ & $.241$ & $.728$ & $.031$ & $.000$ & $%
.965$ & $.035$ \\ \hline
& $200$ & $.492$ & $.371$ & $.137$ & $.795$ & $.155$ & $.050$ & $.648$ & $%
.268$ & $.084$ \\
& $500$ & $.227$ & $.640$ & $.133$ & $.829$ & $.135$ & $.036$ & $.525$ & $%
.415$ & $.060$ \\
$.8$ & $1000$ & $.028$ & $.863$ & $.109$ & $.795$ & $.180$ & $.025$ & $.312$
& $.621$ & $.067$ \\
& $2000$ & $.000$ & $.861$ & $.139$ & $.736$ & $.226$ & $.038$ & $.123$ & $%
.810$ & $.067$ \\
& $3000$ & $.000$ & $.867$ & $.133$ & $.689$ & $.270$ & $.041$ & $.047$ & $%
.866$ & $.087$ \\ \hline
\end{tabular}%
$%
\end{center}
\label{table3}
\end{table}

\begin{table}[tbp]
\caption{Relative frequency detection of 0, 1 and more than 1 change points with the VAR(1) model with a single change point for $z_1=.75$
and with a initial nominal significant level of $\alpha_0=0.05$.}
\begin{center}
$%
\begin{tabular}{ccccccccccc}
\cline{3-11}
&  & \multicolumn{3}{c}{$\left( \rho _{0}=.25,\rho _{1}=-.25\right) $} &
\multicolumn{3}{c}{$\left( \rho _{0}=.25,\rho _{1}=.15\right) $} &
\multicolumn{3}{c}{$\left( \rho _{0}=.25,\rho _{1}=.5\right) $} \\
\cline{3-11}
&  & \multicolumn{3}{c}{Rel. freq.} & \multicolumn{3}{c}{Rel. freq.} &
\multicolumn{3}{c}{Rel. freq.} \\ \cline{3-11}
$\phi $ & $T$ & $0$ & $1$ & $\geq 2$ & $0$ & $1$ & $\geq 2$ & $0$ & $1$ & $%
\geq 2$ \\ \hline
& $200$ & $.634$ & $.355$ & $.011$ & $.954$ & $.045$ & $.001$ & $.820$ & $%
.177$ & $.003$ \\
& $500$ & $.177$ & $.784$ & $.039$ & $.921$ & $.076$ & $.003$ & $.663$ & $%
.324$ & $.013$ \\
$-.5$ & $1000$ & $.005$ & $.950$ & $.045$ & $.865$ & $.130$ & $.005$ & $.318$
& $.652$ & $.030$ \\
& $2000$ & $.000$ & $.944$ & $.056$ & $.771$ & $.219$ & $.010$ & $.054$ & $%
.911$ & $.035$ \\
& $3000$ & $.000$ & $.926$ & $.074$ & $.694$ & $.291$ & $.015$ & $.006$ & $%
.940$ & $.054$ \\ \hline
& $200$ & $.445$ & $.540$ & $.015$ & $.946$ & $.052$ & $.002$ & $.812$ & $%
.187$ & $.001$ \\
& $500$ & $.034$ & $.928$ & $.038$ & $.918$ & $.080$ & $.002$ & $.491$ & $%
.502$ & $.007$ \\
$0$ & $1000$ & $.000$ & $.955$ & $.045$ & $.839$ & $.159$ & $.002$ & $.138$
& $.837$ & $.025$ \\
& $2000$ & $.000$ & $.942$ & $.058$ & $.668$ & $.317$ & $.015$ & $.006$ & $%
.957$ & $.037$ \\
& $3000$ & $.000$ & $.947$ & $.053$ & $.502$ & $.476$ & $.022$ & $.000$ & $%
.949$ & $.051$ \\ \hline
& $200$ & $.635$ & $.262$ & $.103$ & $.806$ & $.143$ & $.051$ & $.739$ & $%
.192$ & $.069$ \\
& $500$ & $.448$ & $.458$ & $.094$ & $.838$ & $.143$ & $.019$ & $.674$ & $%
.292$ & $.034$ \\
$.8$ & $1000$ & $.147$ & $.734$ & $.119$ & $.822$ & $.147$ & $.031$ & $.505$
& $.450$ & $.045$ \\
& $2000$ & $.016$ & $.848$ & $.136$ & $.783$ & $.183$ & $.034$ & $.335$ & $%
.601$ & $.064$ \\
& $3000$ & $.000$ & $.834$ & $.166$ & $.771$ & $.203$ & $.026$ & $.168$ & $%
.738$ & $.094$ \\ \hline
\end{tabular}%
$%
\end{center}
\label{table4}
\end{table}

\begin{sidewaystable}[tbp]
\caption{Median and MAD of the change point estimators for the results in Tables 2, 3 and 4.}
\begin{center}
$%
\begin{tabular}{ccccccccccc}
\cline{3-11}
&  & \multicolumn{3}{c}{$z_{1}=0.25$} & \multicolumn{3}{c}{$z_{2}=0.5$} &
\multicolumn{3}{c}{$z_{3}=0.75$} \\ \cline{3-11}
&  & $\left( \rho _{0},\rho _{1}\right) $ & $\left( \rho _{0},\rho
_{1}\right) $ & $\left( \rho _{0},\rho _{1}\right) $ & $\left( \rho
_{0},\rho _{1}\right) $ & $\left( \rho _{0},\rho _{1}\right) $ & $\left(
\rho _{0},\rho _{1}\right) $ & $\left( \rho _{0},\rho _{1}\right) $ & $%
\left( \rho _{0},\rho _{1}\right) $ & $\left( \rho _{0},\rho _{1}\right) $
\\ \cline{3-11}
&  & $\left( .25,-.25\right) $ & $\left( .25,.15\right) $ & $\left(
.25,.5\right) $ & $\left( .25,-.25\right) $ & $\left( .25,.15\right) $ & $%
\left( .25,.5\right) $ & $\left( .25,-.25\right) $ & $\left( .25,.15\right) $
& $\left( .25,.5\right) $ \\ \cline{3-11}
$\phi $ & $T$ & $\widehat{z}_{1}$ & $\widehat{z}_{1}$ & $\widehat{z}_{1}$ & $%
\widehat{z}_{1}$ & $\widehat{z}_{1}$ & $\widehat{z}_{1}$ & $\widehat{z}_{1}$
& $\widehat{z}_{1}$ & $\widehat{z}_{1}$ \\ \hline
& $200$ & $\underset{\left( .070\right) }{.330}$ & $\underset{\left(
.077\right) }{.532}$ & $\underset{\left( .075\right) }{.335}$ & $\underset{%
\left( .035\right) }{.515}$ & $\underset{\left( .090\right) }{.540}$ & $%
\underset{\left( .065\right) }{.500}$ & $\underset{\left( .055\right) }{.705}
$ & $\underset{\left( .105\right) }{.570}$ & $\underset{\left( .105\right) }{%
.615}$ \\
& $500$ & $\underset{\left( .038\right) }{.288}$ & $\underset{\left(
.092\right) }{.454}$ & $\underset{\left( .068\right) }{.316}$ & $\underset{%
\left( .020\right) }{.506}$ & $\underset{\left( .076\right) }{.522}$ & $%
\underset{\left( .044\right) }{.500}$ & $\underset{\left( .026\right) }{.734}
$ & $\underset{\left( .121\right) }{.541}$ & $\underset{\left( .077\right) }{%
.675}$ \\
$-.5$ & $1000$ & $\underset{\left( .018\right) }{.268}$ & $\underset{\left(
.114\right) }{.414}$ & $\underset{\left( .044\right) }{.284}$ & $\underset{%
\left( .009\right) }{.502}$ & $\underset{\left( .063\right) }{.506}$ & $%
\underset{\left( .024\right) }{.499}$ & $\underset{\left( .015\right) }{.740}
$ & $\underset{\left( .106\right) }{.618}$ & $\underset{\left( .050\right) }{%
.701}$ \\
& $2000$ & $\underset{\left( .009\right) }{.259}$ & $\underset{\left(
.100\right) }{.376}$ & $\underset{\left( .022\right) }{.266}$ & $\underset{%
\left( .005\right) }{.501}$ & $\underset{\left( .055\right) }{.502}$ & $%
\underset{\left( .015\right) }{.499}$ & $\underset{\left( .008\right) }{.744}
$ & $\underset{\left( .088\right) }{.661}$ & $\underset{\left( .023\right) }{%
.731}$ \\
& $3000$ & $\underset{\left( .006\right) }{.255}$ & $\underset{\left(
.099\right) }{.360}$ & $\underset{\left( .016\right) }{.261}$ & $\underset{%
\left( .004\right) }{.501}$ & $\underset{\left( .038\right) }{.500}$ & $%
\underset{\left( .011\right) }{.500}$ & $\underset{\left( .005\right) }{.746}
$ & $\underset{\left( .072\right) }{.676}$ & $\underset{\left( .017\right) }{%
.737}$ \\ \hline
& $200$ & $\underset{\left( .055\right) }{.310}$ & $\underset{\left(
.102\right) }{.527}$ & $\underset{\left( .090\right) }{.335}$ & $\underset{%
\left( .025\right) }{.505}$ & $\underset{\left( .080\right) }{.545}$ & $%
\underset{\left( .055\right) }{.500}$ & $\underset{\left( .042\right) }{.720}
$ & $\underset{\left( .097\right) }{.555}$ & $\underset{\left( .105\right) }{%
.625}$ \\
& $500$ & $\underset{\left( .024\right) }{.272}$ & $\underset{\left(
.088\right) }{.468}$ & $\underset{\left( .052\right) }{.296}$ & $\underset{%
\left( .012\right) }{.502}$ & $\underset{\left( .072\right) }{.514}$ & $%
\underset{\left( .032\right) }{.500}$ & $\underset{\left( .019\right) }{.737}
$ & $\underset{\left( .115\right) }{.599}$ & $\underset{\left( .056\right) }{%
.694}$ \\
$0$ & $1000$ & $\underset{\left( .013\right) }{.262}$ & $\underset{\left(
.098\right) }{.385}$ & $\underset{\left( .031\right) }{.274}$ & $\underset{%
\left( .006\right) }{.501}$ & $\underset{\left( .056\right) }{.511}$ & $%
\underset{\left( .019\right) }{.499}$ & $\underset{\left( .009\right) }{.744}
$ & $\underset{\left( .095\right) }{.657}$ & $\underset{\left( .030\right) }{%
.725}$ \\
& $2000$ & $\underset{\left( .005\right) }{.255}$ & $\underset{\left(
.072\right) }{.327}$ & $\underset{\left( .015\right) }{.260}$ & $\underset{%
\left( .003\right) }{.500}$ & $\underset{\left( .040\right) }{.504}$ & $%
\underset{\left( .009\right) }{.500}$ & $\underset{\left( .005\right) }{.747}
$ & $\underset{\left( .066\right) }{.686}$ & $\underset{\left( .016\right) }{%
.736}$ \\
& $3000$ & $\underset{\left( .004\right) }{.254}$ & $\underset{\left(
.056\right) }{.312}$ & $\underset{\left( .011\right) }{.258}$ & $\underset{%
\left( .002\right) }{.500}$ & $\underset{\left( .034\right) }{.500}$ & $%
\underset{\left( .006\right) }{.500}$ & $\underset{\left( .003\right) }{.747}
$ & $\underset{\left( .056\right) }{.697}$ & $\underset{\left( .012\right) }{%
.740}$ \\ \hline
& $200$ & $\underset{\left( .120\right) }{.395}$ & $\underset{\left(
.120\right) }{.515}$ & $\underset{\left( .140\right) }{.405}$ & $\underset{%
\left( .060\right) }{.520}$ & $\underset{\left( .125\right) }{.535}$ & $%
\underset{\left( .100\right) }{.500}$ & $\underset{\left( .102\right) }{.667}
$ & $\underset{\left( .120\right) }{.545}$ & $\underset{\left( .137\right) }{%
.572}$ \\
& $500$ & $\underset{\left( .078\right) }{.338}$ & $\underset{\left(
.106\right) }{.482}$ & $\underset{\left( .100\right) }{.354}$ & $\underset{%
\left( .041\right) }{.514}$ & $\underset{\left( .096\right) }{.538}$ & $%
\underset{\left( .078\right) }{.500}$ & $\underset{\left( .058\right) }{.710}
$ & $\underset{\left( .098\right) }{.572}$ & $\underset{\left( .121\right) }{%
.595}$ \\
$.8$ & $1000$ & $\underset{\left( .044\right) }{.294}$ & $\underset{\left(
.111\right) }{.459}$ & $\underset{\left( .078\right) }{.322}$ & $\underset{%
\left( .022\right) }{.505}$ & $\underset{\left( .092\right) }{.515}$ & $%
\underset{\left( .060\right) }{.504}$ & $\underset{\left( .031\right) }{.729}
$ & $\underset{\left( .137\right) }{.579}$ & $\underset{\left( .109\right) }{%
.633}$ \\
& $2000$ & $\underset{\left( .022\right) }{.271}$ & $\underset{\left(
.118\right) }{.451}$ & $\underset{\left( .052\right) }{.292}$ & $\underset{%
\left( .135\right) }{.503}$ & $\underset{\left( .082\right) }{.510}$ & $%
\underset{\left( .034\right) }{.500}$ & $\underset{\left( .021\right) }{.737}
$ & $\underset{\left( .120\right) }{.594}$ & $\underset{\left( .056\right) }{%
.697}$ \\
& $3000$ & $\underset{\left( .016\right) }{.267}$ & $\underset{\left(
.122\right) }{.412}$ & $\underset{\left( .039\right) }{.282}$ & $\underset{%
\left( .009\right) }{.502}$ & $\underset{\left( .062\right) }{.500}$ & $%
\underset{\left( .025\right) }{.501}$ & $\underset{\left( .014\right) }{.741}
$ & $\underset{\left( .112\right) }{.627}$ & $\underset{\left( .050\right) }{%
.702}$ \\ \hline
\end{tabular}%
$%
\end{center}
\label{table5}
\end{sidewaystable}

\begin{sidewaystable}[tbp]
\caption{Relative frequency detection of 0, 1, 2 and more than 2 change points with the VAR(1) model with two change points
and with a initial nominal significant level of $\alpha_0=0.05$.}
\begin{center}
$%
\begin{tabular}{cccccccccccccccc}
\cline{5-16}
&  &  &  & \multicolumn{4}{c}{$\phi =-.5$} & \multicolumn{4}{c}{$\phi =0$} &
\multicolumn{4}{c}{$\phi =0.8$} \\ \cline{5-16}
&  &  &  & \multicolumn{4}{c}{$\left( z_{1},z_{2}\right) =\left(
.25,.75\right) $} & \multicolumn{4}{c}{$\left( z_{1},z_{2}\right) =\left(
.25,.75\right) $} & \multicolumn{4}{c}{$\left( z_{1},z_{2}\right) =\left(
.25,.75\right) $} \\ \cline{5-16}
&  &  &  & \multicolumn{4}{c}{Rel. freq.} & \multicolumn{4}{c}{Rel. freq.} &
\multicolumn{4}{c}{Rel. freq.} \\ \cline{5-16}
$\rho _{0}$ & $\rho _{1}$ & $\rho _{2}$ & $T$ & $0$ & $1$ & $2$ & $\geq 3$ &
$0$ & $1$ & $2$ & $\geq 3$ & $0$ & $1$ & $2$ & $\geq 3$ \\ \hline
&  &  & $200$ & $.860$ & $.100$ & $.038$ & $.002$ & $.819$ & $.122$ & $.059$
& $.000$ & $.747$ & $.161$ & $.056$ & $.036$ \\
&  &  & $500$ & $.514$ & $.092$ & $.382$ & $.012$ & $.292$ & $.043$ & $.654$
& $.011$ & $.631$ & $.158$ & $.187$ & $.024$ \\
$.25$ & $-.25$ & $.25$ & $1000$ & $.058$ & $.014$ & $.891$ & $.037$ & $.000$
& $.001$ & $.974$ & $.025$ & $.327$ & $.116$ & $.485$ & $.072$ \\
&  &  & $2000$ & $.000$ & $.000$ & $.949$ & $.051$ & $.000$ & $.000$ & $.963$
& $.037$ & $.046$ & $.025$ & $.814$ & $.115$ \\
&  &  & $3000$ & $.000$ & $.000$ & $.953$ & $.047$ & $.000$ & $.000$ & $.950$
& $.050$ & $.003$ & $.003$ & $.864$ & $.130$ \\ \hline
&  &  & $200$ & $.778$ & $.176$ & $.045$ & $.001$ & $.669$ & $.262$ & $.068$
& $.001$ & $.702$ & $.197$ & $.079$ & $.022$ \\
&  &  & $500$ & $.354$ & $.359$ & $.283$ & $.004$ & $.136$ & $.396$ & $.462$
& $.006$ & $.576$ & $.240$ & $.160$ & $.024$ \\
$.25$ & $.5$ & $0$ & $1000$ & $.025$ & $.297$ & $.662$ & $.016$ & $.000$ & $%
.130$ & $.845$ & $.025$ & $.261$ & $.340$ & $.364$ & $.035$ \\
&  &  & $2000$ & $.000$ & $.059$ & $.902$ & $.039$ & $.000$ & $.004$ & $.973$
& $.023$ & $.036$ & $.230$ & $.656$ & $.078$ \\
&  &  & $3000$ & $.000$ & $.009$ & $.954$ & $.037$ & $.000$ & $.000$ & $.962$
& $.038$ & $.003$ & $.124$ & $.777$ & $.096$ \\ \hline
&  &  & $200$ & $.945$ & $.047$ & $.007$ & $.001$ & $.934$ & $.058$ & $.008$
& $.000$ & $.800$ & $.138$ & $.046$ & $.016$ \\
&  &  & $500$ & $.870$ & $.094$ & $.036$ & $.000$ & $.804$ & $.127$ & $.069$
& $.000$ & $.790$ & $.149$ & $.050$ & $.011$ \\
$.25$ & $0$ & $.25$ & $1000$ & $.673$ & $.166$ & $.157$ & $.004$ & $.542$ & $%
.115$ & $.340$ & $.003$ & $.741$ & $.158$ & $.092$ & $.009$ \\
&  &  & $2000$ & $.284$ & $.130$ & $.572$ & $.014$ & $.125$ & $.033$ & $.823$
& $.019$ & $.566$ & $.161$ & $.245$ & $.028$ \\
&  &  & $3000$ & $.119$ & $.050$ & $.815$ & $.016$ & $.011$ & $.002$ & $.958$
& $.029$ & $.419$ & $.150$ & $.384$ & $.047$ \\ \hline
\end{tabular}%
$%
\end{center}
\label{table6}
\end{sidewaystable}

\begin{table}[tbp]
\caption{Median and MAD of the change point estimators for the results in Table 6.}
\begin{center}
$%
\begin{tabular}{ccccccc}
\cline{5-7}
&  &  &  & $\phi =-.5$ & $\phi =0$ & $\phi =.8$ \\ \cline{5-7}
&  &  &  & $\left( z_{1},z_{2}\right) =\left( .25,.75\right) $ & $\left(
z_{1},z_{2}\right) =\left( .25,.75\right) $ & $\left( z_{1},z_{2}\right)
=\left( .25,.75\right) $ \\ \cline{5-7}
$\rho _{0}$ & $\rho _{1}$ & $\rho _{2}$ & $T$ & $\left( \widehat{z}_{1},%
\widehat{z}_{2}\right) $ & $\left( \widehat{z}_{1},\widehat{z}_{2}\right) $
& $\left( \widehat{z}_{1},\widehat{z}_{2}\right) $ \\ \hline
&  &  & $200$ & $\left( \underset{.049}{.292},\underset{.017}{.745}\right) $
& $\left( \underset{.045}{.290},\underset{.020}{.745}\right) $ & $\left(
\underset{.087}{.307},\underset{.082}{.702}\right) $ \\
&  &  & $500$ & $\left( \underset{.023}{.271},\underset{.016}{.744}\right) $
& $\left( \underset{.014}{.262},\underset{.012}{.748}\right) $ & $\left(
\underset{.044}{.284},\underset{.040}{.744}\right) $ \\
$.25$ & $-.25$ & $.25$ & $1000$ & $\left( \underset{.012}{.259},\underset{%
.011}{.748}\right) $ & $\left( \underset{.009}{.257},\underset{.008}{.749}%
\right) $ & $\left( \underset{.027}{.276},\underset{.025}{.742}\right) $ \\
&  &  & $2000$ & $\left( \underset{.006}{.255},\underset{.005}{.749}\right) $
& $\left( \underset{.005}{.254},\underset{.003}{.749}\right) $ & $\left(
\underset{.017}{.262},\underset{.013}{.747}\right) $ \\
&  &  & $3000$ & $\left( \underset{.005}{.254},\underset{.003}{.749}\right) $
& $\left( \underset{.002}{.252},\underset{.002}{.749}\right) $ & $\left(
\underset{.012}{.259},\underset{.009}{.747}\right) $ \\ \hline
&  &  & $200$ & $\left( \underset{.070}{.305},\underset{.025}{.750}\right) $
& $\left( \underset{.047}{.272},\underset{.032}{.750}\right) $ & $\left(
\underset{.135}{.370},\underset{.075}{.785}\right) $ \\
&  &  & $500$ & $\left( \underset{.048}{.278},\underset{.016}{.750}\right) $
& $\left( \underset{.034}{.268},\underset{.010}{.750}\right) $ & $\left(
\underset{.067}{.296},\underset{.033}{.749}\right) $ \\
$.25$ & $.5$ & $0$ & $1000$ & $\left( \underset{.027}{.266},\underset{.008}{%
.750}\right) $ & $\left( \underset{.020}{.261},\underset{.005}{.750}\right) $
& $\left( \underset{.057}{.288},\underset{.017}{.751}\right) $ \\
&  &  & $2000$ & $\left( \underset{.018}{.259},\underset{.004}{.750}\right) $
& $\left( \underset{.013}{.256},\underset{.002}{.750}\right) $ & $\left(
\underset{.042}{.274},\underset{.009}{.751}\right) $ \\
&  &  & $3000$ & $\left( \underset{.012}{.256},\underset{.002}{.750}\right) $
& $\left( \underset{.007}{.253},\underset{.002}{.749}\right) $ & $\left(
\underset{.029}{.266},\underset{.007}{.750}\right) $ \\ \hline
&  &  & $200$ & $\left( \underset{.025}{.275},\underset{.070}{.700}\right) $
& $\left( \underset{.125}{.270},\underset{.125}{.677}\right) $ & $\left(
\underset{.170}{.332},\underset{.172}{.650}\right) $ \\
&  &  & $500$ & $\left( \underset{.052}{.296},\underset{.068}{.700}\right) $
& $\left( \underset{.044}{.302},\underset{.034}{.734}\right) $ & $\left(
\underset{.094}{.311},\underset{.110}{.702}\right) $ \\
$.25$ & $0$ & $.25$ & $1000$ & $\left( \underset{.034}{.285},\underset{.027}{%
.735}\right) $ & $\left( \underset{.024}{.270},\underset{.023}{.741}\right) $
& $\left( \underset{.062}{.308},\underset{.064}{.717}\right) $ \\
&  &  & $2000$ & $\left( \underset{.021}{.265},\underset{.019}{.740}\right) $
& $\left( \underset{.013}{.259},\underset{.013}{.743}\right) $ & $\left(
\underset{.031}{.277},\underset{.036}{.726}\right) $ \\
&  &  & $3000$ & $\left( \underset{.016}{.260},\underset{.014}{.742}\right) $
& $\left( \underset{.009}{.255},\underset{.009}{.746}\right) $ & $\left(
\underset{.031}{.277},\underset{.031}{.734}\right) $ \\ \hline
\end{tabular}
$%
\end{center}
\label{table7}
\end{table}

\begin{sidewaystable}[tbp]
\caption{Relative frequency detection of 0, 1, 2 and more than 2 change points with the VAR(1) model with two change points
in the mean and the correlation and with a initial nominal significant level of $\alpha_0=0.05$.}
\begin{center}
$%
\begin{tabular}{cccccccccccccccc}
\cline{5-16}
&  &  &  & \multicolumn{4}{c}{$\phi =-.5$} & \multicolumn{4}{c}{$\phi =0$} &
\multicolumn{4}{c}{$\phi =0.8$} \\ \cline{5-16}
&  &  &  & \multicolumn{4}{c}{$\left( z_{1},z_{2}\right) =\left(
.25,.75\right) $} & \multicolumn{4}{c}{$\left( z_{1},z_{2}\right) =\left(
.25,.75\right) $} & \multicolumn{4}{c}{$\left( z_{1},z_{2}\right) =\left(
.25,.75\right) $} \\ \cline{5-16}
&  &  &  & \multicolumn{4}{c}{Rel. freq.} & \multicolumn{4}{c}{Rel. freq.} &
\multicolumn{4}{c}{Rel. freq.} \\ \cline{5-16}
$\rho _{0}$ & $\rho _{1}$ & $\rho _{2}$ & $T$ & $0$ & $1$ & $2$ & $\geq 3$ &
$0$ & $1$ & $2$ & $\geq 3$ & $0$ & $1$ & $2$ & $\geq 3$ \\ \hline
&  &  & $200$ & $.880$ & $.088$ & $.030$ & $.002$ & $.821$ & $.092$ & $.083$
& $.004$ & $.767$ & $.159$ & $.050$ & $.024$ \\
&  &  & $500$ & $.528$ & $.084$ & $.380$ & $.008$ & $.339$ & $.029$ & $.621$
& $.011$ & $.659$ & $.161$ & $.150$ & $.030$ \\
$.25$ & $-.25$ & $.25$ & $1000$ & $.055$ & $.012$ & $.894$ & $.039$ & $.002$
& $.000$ & $.969$ & $.029$ & $.359$ & $.110$ & $.461$ & $.070$ \\
&  &  & $2000$ & $.000$ & $.000$ & $.961$ & $.039$ & $.000$ & $.000$ & $.963$
& $.037$ & $.073$ & $.019$ & $.779$ & $.129$ \\
&  &  & $3000$ & $.000$ & $.000$ & $.941$ & $.059$ & $.000$ & $.000$ & $.967$
& $.033$ & $.016$ & $.003$ & $.849$ & $.132$ \\ \hline
&  &  & $200$ & $.762$ & $.185$ & $.052$ & $.001$ & $.654$ & $.234$ & $.112$
& $.000$ & $.702$ & $.198$ & $.067$ & $.033$ \\
&  &  & $500$ & $.351$ & $.297$ & $.346$ & $.006$ & $.158$ & $.217$ & $.612$
& $.013$ & $.546$ & $.250$ & $.172$ & $.032$ \\
$.25$ & $.5$ & $0$ & $1000$ & $.038$ & $.157$ & $.785$ & $.020$ & $.003$ & $%
.029$ & $.942$ & $.026$ & $.283$ & $.286$ & $.378$ & $.053$ \\
&  &  & $2000$ & $.000$ & $.013$ & $.950$ & $.037$ & $.000$ & $.000$ & $.957$
& $.043$ & $.053$ & $.220$ & $.642$ & $.085$ \\
&  &  & $3000$ & $.000$ & $.001$ & $.946$ & $.053$ & $.000$ & $.000$ & $.967$
& $.033$ & $.002$ & $.115$ & $.790$ & $.093$ \\ \hline
&  &  & $200$ & $.934$ & $.057$ & $.009$ & $.000$ & $.941$ & $.054$ & $.005$
& $.000$ & $.809$ & $.141$ & $.041$ & $.009$ \\
&  &  & $500$ & $.888$ & $.077$ & $.035$ & $.000$ & $.857$ & $.095$ & $.043$
& $.005$ & $.837$ & $.120$ & $.037$ & $.006$ \\
$.25$ & $0$ & $.25$ & $1000$ & $.728$ & $.123$ & $.146$ & $.003$ & $.611$ & $%
.154$ & $.230$ & $.005$ & $.724$ & $.163$ & $.097$ & $.016$ \\
&  &  & $2000$ & $.420$ & $.117$ & $.454$ & $.009$ & $.191$ & $.097$ & $.698$
& $.014$ & $.604$ & $.155$ & $.215$ & $.026$ \\
&  &  & $3000$ & $.162$ & $.063$ & $.742$ & $.033$ & $.032$ & $.020$ & $.918$
& $.030$ & $.477$ & $.124$ & $.359$ & $.040$ \\ \hline
\end{tabular}%
$%
\end{center}
\label{table8}
\end{sidewaystable}

\begin{table}[tbp]
\caption{Median and MAD of the change point estimators for the results in Table 8.}
\begin{center}
$%
\begin{tabular}{ccccccc}
\cline{5-7}
&  &  &  & $\phi =-.5$ & $\phi =0$ & $\phi =.8$ \\ \cline{5-7}
&  &  &  & $\left( z_{1},z_{2}\right) =\left( .25,.75\right) $ & $\left(
z_{1},z_{2}\right) =\left( .25,.75\right) $ & $\left( z_{1},z_{2}\right)
=\left( .25,.75\right) $ \\ \cline{5-7}
$\rho _{0}$ & $\rho _{1}$ & $\rho _{2}$ & $T$ & $\left( \widehat{z}_{1},%
\widehat{z}_{2}\right) $ & $\left( \widehat{z}_{1},\widehat{z}_{2}\right) $
& $\left( \widehat{z}_{1},\widehat{z}_{2}\right) $ \\ \hline
&  &  & $200$ & $\left( \underset{.030}{.300},\underset{.032}{.735}\right) $
& $\left( \underset{.030}{.270},\underset{.025}{.750}\right) $ & $\left(
\underset{.080}{.315},\underset{.075}{.745}\right) $ \\
&  &  & $500$ & $\left( \underset{.026}{.274},\underset{.014}{.746}\right) $
& $\left( \underset{.018}{.264},\underset{.010}{.748}\right) $ & $\left(
\underset{.059}{.325},\underset{.032}{.750}\right) $ \\
$.25$ & $-.25$ & $.25$ & $1000$ & $\left( \underset{.016}{.265},\underset{%
.009}{.748}\right) $ & $\left( \underset{.012}{.261},\underset{.006}{.749}%
\right) $ & $\left( \underset{.032}{.280},\underset{.021}{.749}\right) $ \\
&  &  & $2000$ & $\left( \underset{.008}{.257},\underset{.005}{.748}\right) $
& $\left( \underset{.006}{.255},\underset{.003}{.749}\right) $ & $\left(
\underset{.024}{.272},\underset{.010}{.749}\right) $ \\
&  &  & $3000$ & $\left( \underset{.006}{.255},\underset{.003}{.749}\right) $
& $\left( \underset{.004}{.253},\underset{.002}{.749}\right) $ & $\left(
\underset{.016}{.265},\underset{.007}{.748}\right) $ \\ \hline
&  &  & $200$ & $\left( \underset{.030}{.255},\underset{.022}{.752}\right) $
& $\left( \underset{.037}{.262},\underset{.020}{.750}\right) $ & $\left(
\underset{.090}{.315},\underset{.090}{.765}\right) $ \\
&  &  & $500$ & $\left( \underset{.020}{.254},\underset{.022}{.748}\right) $
& $\left( \underset{.016}{.252},\underset{.018}{.750}\right) $ & $\left(
\underset{.065}{.281},\underset{.031}{.750}\right) $ \\
$.25$ & $.5$ & $0$ & $1000$ & $\left( \underset{.015}{.251},\underset{.011}{%
.749}\right) $ & $\left( \underset{.008}{.250},\underset{.009}{.750}\right) $
& $\left( \underset{.047}{.280},\underset{.018}{.751}\right) $ \\
&  &  & $2000$ & $\left( \underset{.006}{.250},\underset{.006}{.750}\right) $
& $\left( \underset{.004}{.250},\underset{.004}{.750}\right) $ & $\left(
\underset{.032}{.266},\underset{.012}{.750}\right) $ \\
&  &  & $3000$ & $\left( \underset{.004}{.250},\underset{.004}{.750}\right) $
& $\left( \underset{.002}{.250},\underset{.003}{.750}\right) $ & $\left(
\underset{.022}{.260},\underset{.008}{.750}\right) $ \\ \hline
&  &  & $200$ & $\left( \underset{.050}{.410},\underset{.035}{.735}\right) $
& $\left( \underset{.040}{.300},\underset{.025}{.745}\right) $ & $\left(
\underset{.155}{.370},\underset{.125}{.755}\right) $ \\
&  &  & $500$ & $\left( \underset{.040}{.316},\underset{.028}{.738}\right) $
& $\left( \underset{.044}{.326},\underset{.020}{.740}\right) $ & $\left(
\underset{.120}{.408},\underset{.050}{.746}\right) $ \\
$.25$ & $0$ & $.25$ & $1000$ & $\left( \underset{.051}{.316},\underset{.017}{%
.741}\right) $ & $\left( \underset{.046}{.321},\underset{.010}{.746}\right) $
& $\left( \underset{.064}{.334},\underset{.059}{.715}\right) $ \\
&  &  & $2000$ & $\left( \underset{.046}{.312},\underset{.009}{.745}\right) $
& $\left( \underset{.049}{.325},\underset{.005}{.747}\right) $ & $\left(
\underset{.052}{.306},\underset{.033}{.728}\right) $ \\
&  &  & $3000$ & $\left( \underset{.044}{.304},\underset{.006}{.745}\right) $
& $\left( \underset{.040}{.318},\underset{.003}{.748}\right) $ & $\left(
\underset{.043}{.289},\underset{.021}{.742}\right) $ \\ \hline
\end{tabular}%
$%
\end{center}
\label{table9}
\end{table}

\begin{table}[tbp]
\caption{Type I errors with the DCC model with a initial nominal significant level of $\alpha_0=0.05$.}
\begin{center}
$%
\begin{tabular}{ccccccc}
\cline{2-7}
& \multicolumn{2}{c}{$\rho _{0}=0$} & \multicolumn{2}{c}{$\rho _{0}=.5$} &
\multicolumn{2}{c}{$\rho _{0}=.8$} \\ \cline{2-7}
& \multicolumn{2}{c}{Rel. freq.} & \multicolumn{2}{c}{Rel. freq.} &
\multicolumn{2}{c}{Rel. freq.} \\ \hline
$T$ & $0$ & $\geq 1$ & $0$ & $\geq 1$ & $0$ & $\geq 1$ \\ \hline
$500$ & $.941$ & $.059$ & $.940$ & $.060$ & $.939$ & $.061$ \\
$1000$ & $.952$ & $.048$ & $.950$ & $.050$ & $.948$ & $.052$ \\
$2000$ & $.962$ & $.038$ & $.934$ & $.066$ & $.927$ & $.073$ \\
$3000$ & $.951$ & $.049$ & $.942$ & $.058$ & $.936$ & $.064$ \\
$4000$ & $.951$ & $.049$ & $.941$ & $.059$ & $.940$ & $.060$ \\ \hline
\end{tabular}%
$%
\end{center}
\label{table10}
\end{table}

\begin{table}[tbp]
\caption{Relative frequency detection of 0, 1 and more than 1 change points with the DCC model with a single change point
and with a initial nominal significant level of $\alpha_0=0.05$.}
\begin{center}
$%
\begin{tabular}{ccccccccccc}
\cline{3-11}
&  & \multicolumn{3}{c}{$\left( \rho _{0}=.5,\rho _{1}=.6\right) $} &
\multicolumn{3}{c}{$\left( \rho _{0}=.5,\rho _{1}=.7\right) $} &
\multicolumn{3}{c}{$\left( \rho _{0}=.5,\rho _{1}=.8\right) $} \\
\cline{3-11}
&  & \multicolumn{3}{c}{Rel. freq.} & \multicolumn{3}{c}{Rel. freq.} &
\multicolumn{3}{c}{Rel. freq.} \\ \cline{3-11}
$z_{1}$ & $T$ & $0$ & $1$ & $\geq 2$ & $0$ & $1$ & $\geq 2$ & $0$ & $1$ & $%
\geq 2$ \\ \hline
& $500$ & $.822$ & $.176$ & $.002$ & $.377$ & $.616$ & $.007$ & $.046$ & $%
.917$ & $.037$ \\
& $1000$ & $.706$ & $.285$ & $.009$ & $.112$ & $.862$ & $.026$ & $.005$ & $%
.939$ & $.056$ \\
$.25$ & $2000$ & $.493$ & $.500$ & $.007$ & $.009$ & $.951$ & $.040$ & $.000$
& $.925$ & $.075$ \\
& $3000$ & $.298$ & $.683$ & $.019$ & $.000$ & $.953$ & $.047$ & $.000$ & $%
.939$ & $.061$ \\
& $4000$ & $.210$ & $.765$ & $.025$ & $.000$ & $.953$ & $.047$ & $.000$ & $%
.941$ & $.059$ \\ \hline
& $500$ & $.737$ & $.260$ & $.003$ & $.319$ & $.667$ & $.014$ & $.047$ & $%
.921$ & $.032$ \\
& $1000$ & $.596$ & $.398$ & $.006$ & $.068$ & $.910$ & $.022$ & $.001$ & $%
.947$ & $.052$ \\
$.5$ & $2000$ & $.299$ & $.685$ & $.016$ & $.003$ & $.961$ & $.036$ & $.000$
& $.951$ & $.049$ \\
& $3000$ & $.155$ & $.825$ & $.020$ & $.000$ & $.965$ & $.035$ & $.000$ & $%
.927$ & $.073$ \\
& $4000$ & $.072$ & $.889$ & $.039$ & $.000$ & $.943$ & $.057$ & $.000$ & $%
.940$ & $.060$ \\ \hline
& $500$ & $.838$ & $.161$ & $.001$ & $.597$ & $.394$ & $.009$ & $.268$ & $%
.706$ & $.026$ \\
& $1000$ & $.790$ & $.203$ & $.007$ & $.279$ & $.700$ & $.021$ & $.031$ & $%
.933$ & $.036$ \\
$.75$ & $2000$ & $.592$ & $.399$ & $.009$ & $.050$ & $.923$ & $.027$ & $.000$
& $.929$ & $.071$ \\
& $3000$ & $.423$ & $.563$ & $.014$ & $.002$ & $.957$ & $.041$ & $.000$ & $%
.941$ & $.059$ \\
& $4000$ & $.275$ & $.711$ & $.014$ & $.000$ & $.956$ & $.044$ & $.000$ & $%
.935$ & $.065$ \\ \hline
\end{tabular}%
$%
\end{center}
\label{table11}
\end{table}

\begin{table}[tbp]
\caption{Median and MAD of the change point estimators for the results in Table 11.}
\begin{center}
$%
\begin{tabular}{ccccc}
\cline{3-5}
&  & $\left( \rho _{0},\rho _{1}\right) $ & $\left( \rho _{0},\rho
_{1}\right) $ & $\left( \rho _{0},\rho _{1}\right) $ \\ \cline{3-5}
&  & $\left( .5,.6\right) $ & $\left( .5,.7\right) $ & $\left( .5,.8\right) $
\\ \cline{3-5}
$z_{1}$ & $T$ & $\widehat{z}_{1}$ & $\widehat{z}_{1}$ & $\widehat{z}_{1}$ \\
\hline
& $500$ & $\underset{\left( .121\right) }{.383}$ & $\underset{\left(
.058\right) }{.300}$ & $\underset{\left( .036\right) }{.284}$ \\
& $1000$ & $\underset{\left( .074\right) }{.319}$ & $\underset{\left(
.035\right) }{.279}$ & $\underset{\left( .022\right) }{.272}$ \\
$0.25$ & $2000$ & $\underset{\left( .059\right) }{.301}$ & $\underset{\left(
.021\right) }{.268}$ & $\underset{\left( .014\right) }{.265}$ \\
& $3000$ & $\underset{\left( .045\right) }{.291}$ & $\underset{\left(
.017\right) }{.264}$ & $\underset{\left( .009\right) }{.259}$ \\
& $4000$ & $\underset{\left( .038\right) }{.285}$ & $\underset{\left(
.013\right) }{.262}$ & $\underset{\left( .008\right) }{.257}$ \\ \hline
& $500$ & $\underset{\left( .090\right) }{.491}$ & $\underset{\left(
.040\right) }{.508}$ & $\underset{\left( .024\right) }{.510}$ \\
& $1000$ & $\underset{\left( .059\right) }{.505}$ & $\underset{\left(
.024\right) }{.506}$ & $\underset{\left( .013\right) }{.506}$ \\
$0.5$ & $2000$ & $\underset{\left( .043\right) }{.502}$ & $\underset{\left(
.013\right) }{.502}$ & $\underset{\left( .006\right) }{.503}$ \\
& $3000$ & $\underset{\left( .030\right) }{.500}$ & $\underset{\left(
.010\right) }{.501}$ & $\underset{\left( .004\right) }{.502}$ \\
& $4000$ & $\underset{\left( .026\right) }{.499}$ & $\underset{\left(
.006\right) }{.502}$ & $\underset{\left( .003\right) }{.501}$ \\ \hline
& $500$ & $\underset{\left( .128\right) }{.620}$ & $\underset{\left(
.068\right) }{.698}$ & $\underset{\left( .042\right) }{.720}$ \\
& $1000$ & $\underset{\left( .114\right) }{.626}$ & $\underset{\left(
.045\right) }{.715}$ & $\underset{\left( .020\right) }{.738}$ \\
$0.75$ & $2000$ & $\underset{\left( .073\right) }{.678}$ & $\underset{\left(
.021\right) }{.736}$ & $\underset{\left( .009\right) }{.745}$ \\
& $3000$ & $\underset{\left( .046\right) }{.707}$ & $\underset{\left(
.014\right) }{.739}$ & $\underset{\left( .005\right) }{.747}$ \\
& $4000$ & $\underset{\left( .037\right) }{.718}$ & $\underset{\left(
.011\right) }{.742}$ & $\underset{\left( .004\right) }{.748}$ \\ \hline
\end{tabular}%
$%
\end{center}
\label{table12}
\end{table}

\begin{table}[tbp]
\caption{Relative frequency detection of 0, 1, 2 and more than 2 change points with the DCC model with two change points
and with a initial nominal significant level of $\alpha_0=0.05$.}
\begin{center}
$%
\begin{tabular}{cccccccc}
\cline{5-8}
&  &  &  & \multicolumn{4}{c}{$\left( z_{1},z_{2}\right) =\left(
.25,.75\right) $} \\ \cline{5-8}
&  &  &  & \multicolumn{4}{c}{Rel. freq.} \\ \cline{5-8}
$\rho _{0}$ & $\rho _{1}$ & $\rho _{2}$ & $T$ & $0$ & $1$ & $2$ & $\geq 3$
\\ \hline
&  &  & $500$ & $.716$ & $.170$ & $.108$ & $.006$ \\
&  &  & $1000$ & $.405$ & $.115$ & $.461$ & $.019$ \\
$.5$ & $.7$ & $.5$ & $2000$ & $.073$ & $.019$ & $.864$ & $.044$ \\
&  &  & $3000$ & $.010$ & $.000$ & $.949$ & $.041$ \\
&  &  & $4000$ & $.000$ & $.000$ & $.954$ & $.046$ \\ \hline
&  &  & $500$ & $.597$ & $.363$ & $.040$ & $.000$ \\
&  &  & $1000$ & $.358$ & $.470$ & $.169$ & $.003$ \\
$.5$ & $.7$ & $.6$ & $2000$ & $.081$ & $.352$ & $.543$ & $.024$ \\
&  &  & $3000$ & $.013$ & $.161$ & $.788$ & $.038$ \\
&  &  & $4000$ & $.001$ & $.063$ & $.892$ & $.044$ \\ \hline
&  &  & $500$ & $.575$ & $.423$ & $.002$ & $.000$ \\
&  &  & $1000$ & $.342$ & $.642$ & $.016$ & $.000$ \\
$.5$ & $.6$ & $.7$ & $2000$ & $.080$ & $.842$ & $.076$ & $.002$ \\
&  &  & $3000$ & $.030$ & $.738$ & $.225$ & $.007$ \\
&  &  & $4000$ & $.004$ & $.575$ & $.400$ & $.021$ \\ \hline
\end{tabular}%
$%
\end{center}
\label{table13}
\end{table}

\begin{table}[tbp]
\caption{Median and MAD of the change point estimators for the results in Table 13.}
\begin{center}
$%
\begin{tabular}{ccccc}
\cline{5-5}
&  &  &  & $\left( z_{1},z_{2}\right) =\left( .25,.75\right) $ \\ \cline{5-5}
$\rho _{0}$ & $\rho _{1}$ & $\rho _{2}$ & $T$ & $\left( \widehat{z}_{1},%
\widehat{z}_{2}\right) $ \\ \hline
&  &  & $500$ & $\left( \underset{.051}{.278},\underset{.028}{.750}\right) $
\\
&  &  & $1000$ & $\left( \underset{.029}{.269},\underset{.016}{.751}\right) $
\\
$.5$ & $.7$ & $.5$ & $2000$ & $\left( \underset{.015}{.260},\underset{.010}{%
.750}\right) $ \\
&  &  & $3000$ & $\left( \underset{.011}{.258},\underset{.006}{.750}\right) $
\\
&  &  & $4000$ & $\left( \underset{.007}{.255},\underset{.005}{.750}\right) $
\\ \hline
&  &  & $500$ & $\left( \underset{.056}{.330},\underset{.032}{.747}\right) $
\\
&  &  & $1000$ & $\left( \underset{.033}{.274},\underset{.045}{.737}\right) $
\\
$.5$ & $.7$ & $.6$ & $2000$ & $\left( \underset{.017}{.264},\underset{.026}{%
.743}\right) $ \\
&  &  & $3000$ & $\left( \underset{.012}{.257},\underset{.019}{.743}\right) $
\\
&  &  & $4000$ & $\left( \underset{.009}{.257},\underset{.017}{.743}\right) $
\\ \hline
&  &  & $500$ & $\left( \underset{.019}{.073},\underset{.240}{.638}\right) $
\\
&  &  & $1000$ & $\left( \underset{.042}{.243},\underset{.029}{.762}\right) $
\\
$.5$ & $.6$ & $.7$ & $2000$ & $\left( \underset{.040}{.256},\underset{.027}{%
.764}\right) $ \\
&  &  & $3000$ & $\left( \underset{.026}{.257},\underset{.023}{.751}\right) $
\\
&  &  & $4000$ & $\left( \underset{.023}{.257},\underset{.020}{.750}\right) $
\\ \hline
\end{tabular}%
$%
\end{center}
\label{table14}
\end{table}

\begin{table}[tbp]
\caption{Iterations taken by the procedure in the real data example, (*) means statistically significant change point. The initial nominal significant level is $\alpha_0=0.05$}
\begin{center}
$%
\begin{tabular}{ccccc}
\hline
\multicolumn{5}{c}{Step 1} \\ \hline
Interval & $Q_{T}(X,Y)$ & Change point & Time point & Date \\ \hline
$\left[ 1,3524\right] $ & 1.5700 (*) & 988 & 0.2804 & November 29, 2000 \\
\hline
\multicolumn{5}{c}{Step 2} \\ \hline
Interval & $Q_{T}(X,Y)$ & Change point & Time point & Date \\ \hline
$\left[ 1,988\right] $ & 2.1009 (*) & 664 & 0.1884 & August 19, 1999 \\
$\left[ 989,3524\right] $ & 1.4745 & 2966 & 0.8417 & October 14, 2008 \\ \hline
$\left[ 1,664\right] $ & 1.0482 & 157 & 0.0446 & August 14, 1997 \\
$\left[ 665,988\right] $ & 1.3471 & 825 & 0.2341 & April 7, 2000 \\
%\hline
$\left[ 989,3524\right] $ & 1.4745 & 2966 & 0.8417 & October 14, 2008 \\
\hline
\multicolumn{5}{c}{Step 3} \\ \hline
Interval & $Q_{T}(X,Y)$ & Change point & Time point & Date \\ \hline
$\left[ 1,988\right] $ & 2.1009 (*) & 664 & 0.1884 & August 19, 1999 \\
$\left[ 665,3524\right] $ & 1.6193 (*) & 2734 & 0.7758 & November 12, 2007
\\ \hline
\end{tabular}
$%
\end{center}
\label{table15}
\end{table}

\newpage

\clearpage

\clearpage

\clearpage

\clearpage

\end{document}